\RequirePackage{fix-cm}
\documentclass[smallextended]{svjour3}       
\smartqed  
\usepackage{graphicx}
\usepackage{mathptmx}      
\usepackage{amsmath}      
\newcommand{\teff}{$T_\mathrm{{eff}}$}

\def\spose#1{\hbox to 0pt{#1\hss}}
\def\deg{\ifmmode^\circ\else$\null^\circ$\fi}
\def\gta{\mathrel{\spose{\lower 3pt\hbox{$\mathchar "218$}}\raise 2.0pt\hbox{$\mathchar"13E$}}}
\def\lta{\mathrel{\spose{\lower 3pt\hbox{$\mathchar "218$}}\raise 2.0pt\hbox{$\mathchar"13C$}}}
\def\scinot#1.{\hbox{$\, \times \, 10^{#1}$}}

\journalname{Experimental Astronomy}
\begin{document}

\title{Chemical variation with altitude and longitude on exo-Neptunes: Predictions for Ariel phase-curve observations
\thanks{This work was supported by the National Aeronautics and Space Administration under 
grant number NNX16AC64G issued through the Exoplanets Research Program.}
}

\titlerunning{Chemical variation with altitude and longitude on exo-Neptunes}        

\author{Julianne I. Moses \and
        Pascal Tremblin  \and
	Olivia Venot  \and
	Yamila Miguel
}


\institute{J.~I. Moses \at
              Space Science Institute\\
	      4765 Walnut St, Suite B \\
	      Boulder, CO 80301, USA \\
              \email{jmoses@spacescience.org}   
           \and
           P. Tremblin \at
           Maison de la Simulation\\
	   CEA, CNRS, Univ.~Paris-Sud\\
	   UVSQ, Univ.~Paris-Saclay\\
	   91191 Gif-sur-Yvette, France
           \and
           O. Venot \at
           LISA, UMR CNRS 7583 \\
           Univ.~Paris-Est-Cr{\'e}teil \\
	   Univ.~Paris, Institut Pierre Simon Laplace\\
	   Cr{\'e}teil, France
	   \and
           Y. Miguel \at
           Leiden Observatory\\
           Univ.~Leiden \\
           Niels Bohrweg 2\\
           2333CA Leiden, The Netherlands\\
	   \\
	   SRON Netherlands Institute for Space Research \\
	   Sorbonnelaan 2 \\
	   3584 CA, Utrecht, The Netherlands
}

\date{Received: date / Accepted: date}

\maketitle

\begin{abstract}
Using two-dimensional (2D) thermal structure models and pseudo-2D chemical kinetics models, we explore how atmospheric temperatures and composition 
change as a function of altitude and longitude within the equatorial regions of close-in transiting Neptune-class exoplanets at different distances 
from their host stars.  Our models predict that the day-night stratospheric temperature contrasts increase with increasing planetary effective 
temperatures \teff\ and that the atmospheric composition changes significantly with \teff .  We find that horizontal transport-induced quenching 
is very effective in our simulated exo-Neptune atmospheres, acting to homogenize the vertical profiles of species abundances with longitude at 
stratospheric pressures where infrared observations are sensitive.  
Our models have important implications for planetary emission observations as a function of orbital phase with the \textit{Ariel} mission.  
Cooler solar-composition exo-Neptunes with \teff\ = 500--700 K are strongly affected by photochemistry and other disequilibrium chemical 
processes, but their predicted variations in infrared emission spectra with orbital phase are relatively small, making them less robust 
phase-curve targets for \textit{Ariel} observations.  Hot solar-composition exo-Neptunes with \teff\ $\ge$ 1300 K exhibit strong variations 
in infrared emission with orbital phase, making them great targets for constraining global temperatures, energy-balance details, atmospheric 
dynamics, and the presence of certain high-temperature atmospheric constituents.  However, such high-temperature exo-Neptunes are arguably less 
interesting from an atmospheric chemistry standpoint, with spectral signatures being dominated by a small number of species whose abundances 
are expected to be constant with longitude and consistent with thermochemical equilibrium.  Solar-composition 
exo-Neptunes with \teff\ = 900--1100 K reside in an interesting intermediate regime, with infrared phase curve variations being affected by 
both temperature and composition variations, albeit at smaller predicted phase-curve amplitudes than for the hotter planets.  This interesting 
intermediate regime shifts to smaller temperatures as atmospheric metallicity is increased, making cool higher-metallicity Neptune-class planets 
appropriate targets for \textit{Ariel} phase-curve observations.
\keywords{Exoplanet atmospheres \and Exoplanet atmospheric composition \and Atmospheric chemistry \and Photochemistry \and Thermochemistry}
\end{abstract}

\section{Introduction} \label{sec:intro}

The European Space Agency's \textit{Atmospheric Remote-sensing Infrared Exoplanet Large-survey (Ariel)} mission, which is due to launch in 
2028, is a space-based telescope survey mission dedicated to acquiring simultaneous narrow-band visible photometry and near-infrared spectra of 
$\sim$1000 exoplanets that transit their host stars \cite{tinetti18}.  The mission promises to revolutionize our understanding of planet formation
and atmospheric processes, due to \textit{Ariel}'s ability to uniformly sample and systematically characterize the atmospheres of a statistically 
large sample of exoplanets.  Transit, eclipse, and phase-curve observations and eclipse mapping of exoplanets can provide important information 
on atmospheric temperatures, composition, dynamics, and energy transport \cite{bailey14rev,cowan15,crossfield15,demingseager17,madhu19,madhu16rev}.
The 1.25--7.8 $\mu$m spectral range covered by \textit{Ariel} encompasses infrared molecular bands 
of many key atmospheric species, including H$_2$O, CO, CO$_2$, CH$_4$, NH$_3$, HCN, NO, C$_2$H$_2$, C$_2$H$_6$, PH$_3$, H$_2$S, SiO, TiO, and VO; 
in addition, \textit{Ariel}'s visible-wavelength channels provide complementary information on aerosol extinction.  \textit{Ariel} will focus on 
warm and hot planets, to take advantage of the higher signal-to-noise ratio observations that such planets provide, and to ensure 
minimal sequestering of different elements in condensed phases in their atmospheres compared to solar-system planets, thus allowing a better 
measure of the bulk elemental composition of the atmosphere as a whole \cite{tinetti18}.  Because large H$_2$-rich planets are less likely to have 
lost their primordial atmospheres, the composition and bulk elemental ratios of giant planets, in particular, provide a unique 
record of the origins of planetary systems, supplying clues to how and where the planet formed, whether migration played a role in its current 
location, and how its complement of heavy elements were acquired \cite{helled14lunine,mordasini16}.  The transiting planets that will be observed 
by \textit{Ariel} are exotic by solar-system standards.  They reside close to their stars, experiencing a regime of strong radiative forcing that 
cannot be studied in our own solar system.  \textit{Ariel} will provide information on the current climate, chemistry, physics, and dynamics in the 
atmospheres of these exotic worlds.

Most of the close-in transiting exoplanets that will be observed by \textit{Ariel} are expected to be tidally locked, with one hemisphere constantly 
facing the star.  The permanent day-night radiative forcing from this synchronous rotation leads to globally uneven atmospheric heating, large 
horizontal temperature gradients, and strong atmospheric winds that help redistribute some of the day-side heat to the night side of the planet 
\cite{joshi97,showman10}.  For gas-rich ``hot Neptunes'' and ``hot Jupiters'', one consequence of this strong day-night forcing is a high-velocity 
(up to several km s$^{-1}$) superrotating zonal-wind jet at low latitudes \cite{showman09,lewis10,showman11,zhang17} that can displace the day-side 
radiative hot spot eastward away from the substellar point \cite{showman02,knut07,hammond18,debras20}.  These winds transport atmospheric constituents as 
well as heat.  A low-latitude parcel of gas being carried by these zonal winds will experience an effective day-night cycle, with temperatures and incident 
stellar ultraviolet flux levels that vary as the parcel is transported from the day side to the night side and back again.  The composition of the gas can 
change as a result of this cycle.  Emission observations of a close-in transiting exoplanet over all phases of its orbit around its host star can 
help sample these global variations \cite{cowan11model,komacek17,parmentier18,stevenson14wasp43b,steinrueck19}, furthering our understanding how atmospheric 
dynamics, energy transport, and 
disequilibrium chemistry respond in a three-dimensional sense to the strong day-night forcing.  Such ``phase-curve'' observations consume copious amounts 
of telescope time, however, and upcoming broad-purpose facilities such as the \textit{James Webb Space Telescope (JWST)} will be unable to perform 
full phase-curve observations for many exoplanets \cite{cowan15}.  Being a dedicated exoplanet characterization mission, \textit{Ariel} will be 
better able to accommodate phase-curve observations, which are expected to account for $\sim$10\% of the science time, leading to the acquisition of 
$\sim$35-50 planetary phase curves over \textit{Ariel}'s 3.5-year mission \cite{tinetti18,charnay20}.

Three-dimensional (3D) global climate models that fully couple chemistry, radiative transport, and dynamics are needed to accurately predict the 
observable variations in atmospheric composition and temperature across highly irradiated transiting exoplanets, with the resulting consequences 
for phase-curve observations.  However, the inclusion of full chemical kinetics networks in such models is computationally exorbitant, and intractable 
even with today's fastest super-computers.  Transport-induced quenching --- a disequilibrium chemical process that is likely very important for H$_2$-rich 
exoplanets and brown dwarfs \cite{prinn77,fegley96,saumon00,lodders02,hubeny07,moses11,visscher11,venot12} --- has been included in transiting-planet 
general circulation models (GCMs) \cite{coop06,bordwell18,drummond18,drummond20,mendonca18} via relaxation methods \cite{tsai18relax}, reduced 
chemical reaction networks \cite{venot19reduced}, or other approximations. These simplifications partially resolve the problem, but chemical schemes 
that include photochemistry have not yet been considered in exoplanet GCMs.   Less computationally expensive two-dimensional (2D) models can come to 
the rescue, in this case.  

Ag{\'u}ndez et al. 
\cite{agundez14pseudo,agundez12} describe a novel pseudo-2D model framework that can account for both photochemistry and vertical/horizontal chemical 
quenching.  Their pseudo-2D models take advantage of the fact that strong and stable equatorial zonal jets are predicted to form 
on H$_2$-rich transiting exoplanets, such that their atmospheres experience effective rotation and thus a diurnal cycle.  The zonal jets encompass a broad 
altitude range over the observable region of the planetary atmosphere, and although the jet speed is not expected to be uniform with altitude, the 
assumption that the entire atmospheric column ``rotates'' at a constant rate controlled by the zonal winds provides a useful first-order scenario for 
testing the influence of horizontal chemical quenching and the global survival of photochemically produced species \cite{agundez14pseudo}.  

The pseudo-2D description then reduces to a series of 
time-variable one-dimensional (1D) disequilibrium chemistry calculations that solve for the atmospheric composition as a function of altitude and 
longitude as the physical background conditions change with time/longitude due to variations in the irradiation angle and thermal structure as 
the atmospheric column rotates around the planet.  Vertical transport is considered in these models, but horizontal transport of material in and out 
of the ``rotating'' column is ignored.  As with the 3D GCMs mentioned above that include chemical quenching, the pseudo-2D chemical models 
demonstrate that horizontal chemical quenching is very important on close-in extrasolar giant planets, helping to homogenize the 
chemical composition with longitude \cite{agundez14pseudo,agundez12,venot20wasp43}.
The pseudo-2D description does not perfectly reflect the more complicated and self-consistent 3D behavior \cite{bordwell18,drummond18,drummond20}, 
but the results are much more accurate than assuming thermochemical equilibrium, especially for cooler planets.  Moreover, the warmer low-latitude 
regions where the jets are concentrated dominate the observed emission from the disk of the planet, helping to justify the pseudo-2D approximations for 
predicting phase-curve behavior.

To date, pseudo-2D chemical models have been developed for the hot Jupiters HD 189733 b, HD 209458 b, and WASP-43 b 
\cite{agundez12,agundez14pseudo,venot20wasp43}.  
Here, we expand the pseudo-2D chemical model framework to smaller, more generalized Neptune-class exoplanets (``exo-Neptunes'') at a variety of orbital 
distances from their host stars.
\textit{Kepler} observations indicate that planets of Neptune's size and smaller dominate the exoplanet population in our galaxy 
\cite{fressin13,petigura13,batalha13}, yet we understand little about how such planets form and evolve 
\cite{helled14uran,venturini17,lambrechts19,choksi20}, or how diverse their atmospheres 
could be \cite{kite20,leconte15,mordassini20,moses13gj436}.  As such, exo-Neptunes represent an important class of planets that will be targeted by 
\textit{Ariel}.  

To set up our grid of planets, we use the \texttt{2D-ATMO} steady-state circulation model described by Tremblin et al. \cite{tremblin17} to define the 
equatorial temperature profiles as a function of altitude and longitude on exo-Neptunes at nine different distances from a K5 V star. 
Then, we use a pseudo-2D chemical kinetics model to track how the low-latitude composition of these exoplanets varies as a 
function of altitude and longitude due to the changing incident stellar flux and temperatures as the atmosphere is assumed to rotate as a solid body 
with a rate controlled by the zonal winds.  Both photochemistry and potential vertical and horizontal (zonal) transport-induced quenching are considered in the 
chemical model.  Armed with these pseudo-2D results, we assume that the calculated longitude-dependent composition at the equator is representative of 
all latitudes, and we examine the consequences of this global variation of temperature and composition on the emission spectra of the planets 
as a function of orbital phase.  We compare our pseudo-2D chemical model results with those derived from thermochemical equilibrium, and we examine the 
sensitivity of the results to atmospheric metallicity.  Finally, we discuss the implications for \textit{Ariel} phase-curve observations.

\section{Theoretical models} \label{sec:models}

We consider a grid of Neptune-sized planets, all with gravity $g$ = 1000 cm s$^{-2}$ and radius 0.4 times Jupiter's radius.  These planets are placed 
at various distances from a K5 V star with effective temperature \teff\ = 4500 K, $\log (g)$ = 4.5 (cgs), and a radius 0.7 times that of the Sun.  The 
planets are assumed to have circular orbits with orbital radii that correspond to a planetary \teff\ of 500, 700, 900, 1100, 1300, 1500, 
1700, 1900, and 2100 K, where the planetary $T_\mathrm{{eff}}$ here corresponds to $\sqrt{2}$ times the equilibrium temperature $T_\mathrm{{eq}}$ 
the planet would have for an albedo of zero, an emissivity of 1.0 at all wavelengths, global re-radiation of the incident energy from the star, and no 
internal heat flux, i.e., \teff\ = $T_\mathrm{{star}}$ $\times$ $(R_\mathrm{star}/a)^{1/2}$, where $T_\mathrm{{star}}$ is the stellar effective 
temperature, $R_\mathrm{{star}}$ is the stellar radius, and $a$ is the planetary orbital radius.  The different planetary \teff 's in our above grid 
correspond to, respectively, $a$ = 0.131896, 0.067294, 0.040709, 0.027251, 0.019511, 0.014655, 0.011410, 0.009134, and 0.007477 AU.  

\begin{figure*}[!htb]
\begin{center}
\includegraphics[width=1.0\textwidth]{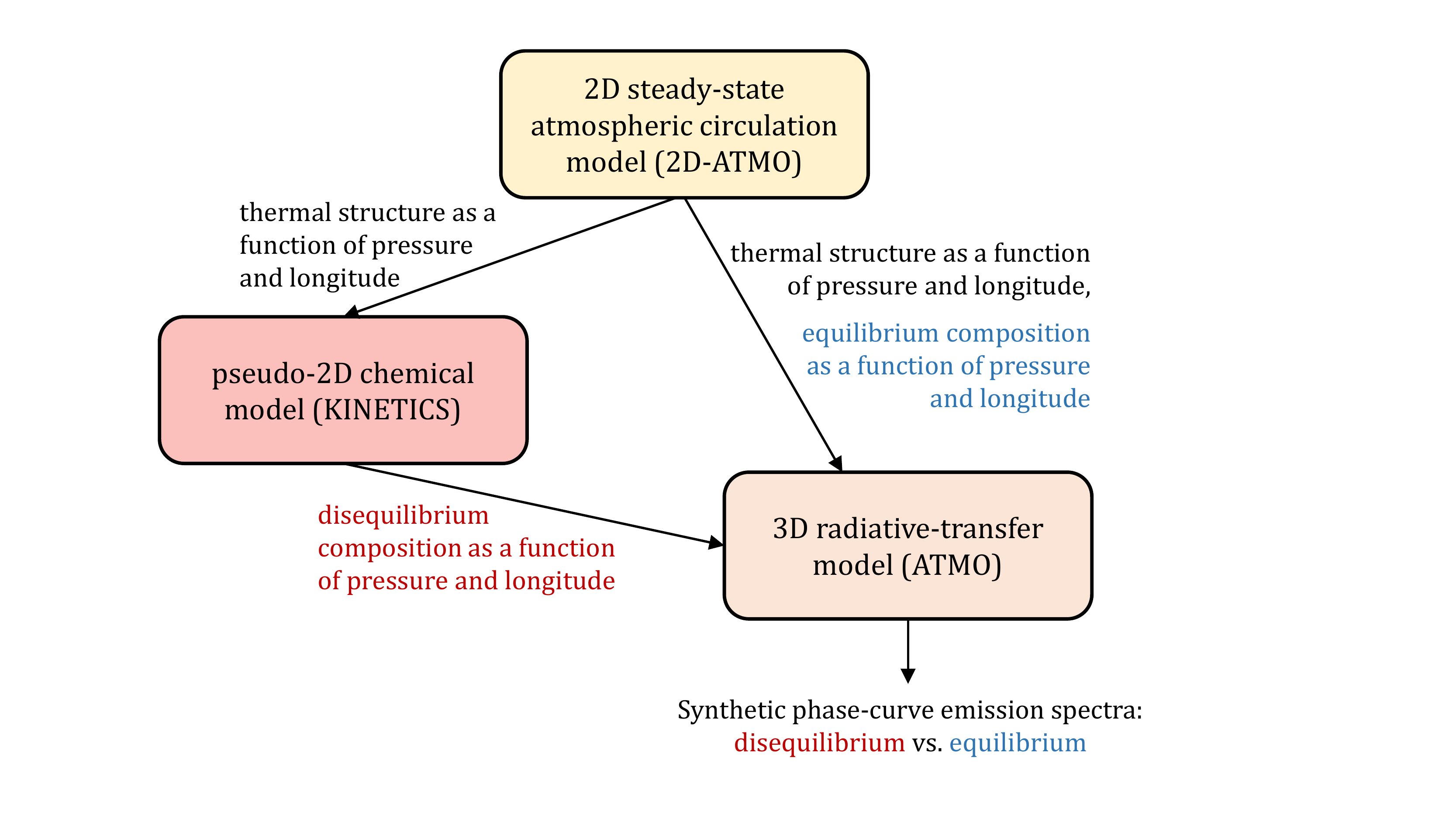}
\end{center}
\vspace{-0.5cm}
\caption{The three main theoretical models used in the study, with the arrows representing the relevant output and links between them 
(see section~\ref{sec:models}).}
\label{figflowchart}       
\end{figure*}

As described below, we use two different theoretical models to predict the thermal structure and chemical composition of these planets, and a third model 
to calculate the resulting emission spectra and consequences for phase-curve observations.  The links and dependencies between the models are shown in 
Fig.~\ref{figflowchart}.

\subsection{2D thermal structure model} \label{sec:2datmo}

The thermal structure as a function of longitude and altitude in the equatorial region of these planets is derived from \texttt{2D-ATMO}, as described in 
Tremblin et al. \cite{tremblin17}.  In this 2D steady-state atmospheric circulation model, the wind is driven by longitudinal pressure gradients, and 
mass is conserved through the longitudinal mass flow being balanced by a combination of the vertical and meridional flow.  The mass fluxes of the meridional and 
vertical components of the wind are assumed to be proportional to each other by a constant $\alpha$.  If $\alpha$ $\rightarrow$ $\infty$, the winds are 
purely longitudinal and meridional; if $\alpha$ $\rightarrow$ 0, the flow is purely longitudinal and vertical.  This constant can be calibrated 
using a 3D GCM, as illustrated in \cite{tremblin17}.  No comparable 3D GCMs were available for our generic Neptune-class planets, so we have simply assumed 
a low internal heat flux and large $\alpha = 10^4$, such that advection of potential temperature to the deep atmosphere is negligible.  We also have assumed 
a constant zonal wind of 4 km s$^{-1}$ at the substellar point for all these models, although one might expect the zonal jet on highly irradiated 
planets to vary with planetary effective temperature (e.g., \cite{showman10}), which can influence zonal temperature variations and the resulting phase 
curves.  See Tremblin et al. \cite{tremblin17} for further information about \texttt{2D-ATMO}.

For the radiative-transfer calculations used in conjunction with the \texttt{2D-ATMO}, the stellar spectrum is adopted from a Kurucz 
model\footnote{http://kurucz.harvard.edu/stars.html} with stellar properties described above in section~\ref{sec:models}.  The planetary atmosphere is 
assumed to be in thermochemical equilibrium, with 277 species considered for the equilibrium calculations; rainout is included for all condensed species 
(see \cite{goyal18,goyal18err}).  
Twenty-two constituents are assumed to contribute to atmospheric opacities: H$_2$O, CO, CO$_2$, CH$_4$, NH$_3$, Na, K, Li, Rb, Cs, TiO, VO, FeH, PH$_3$, 
H$_2$S, HCN, C$_2$H$_2$, SO$_2$, Fe, H$^{-}$, and collision-induced absorption from H$_2$-H$_2$ and H$_2$-He.  Our \texttt{2D-ATMO} calculations consider a grid 
of exo-Neptunes with four different metallicities ---  1$\times$, 10$\times$, 100$\times$, and 1000$\times$ solar metallicity; the opactiy references, 
other model details, and full model output is publicly available.\footnote{http://opendata.erc-atmo.eu}   
In conjunction with the disequilibrium models presented in section \ref{sec:chem}, we focus on the 1$\times$ solar-metallicity results for this paper 
(with the solar composition taken from \cite{caffau11}), but we also examine the results for 100$\times$ solar metallicity for a few planetary \teff\ cases.  
The radiative transfer is solved in plane-parallel geometry and includes isotropic scattering of gases, 
but no absorption or scattering from clouds/hazes.  
The calculations are performed on a grid of 20 longitudes and 51 radius levels; the radius range is the same for all longitudes, such that the pressure 
at the top layer in the model changes with longitude as the atmosphere expands or contracts with underlying temperature changes.  Tremblin et al. \cite{tremblin17} 
demonstrate that 10 longitude grid points are sufficient to recover the dayside-nightside temperature gradient with \texttt{2D-ATMO}; however, we find that a slightly 
higher longitudinal resolution is needed to sample the phase curve correctly.  Increasing beyond 20 longitude points does not further improve the phase-curve 
predictions.  Additional details of the radiative-transfer and chemical-equilibrium procedures can be found elsewhere 
\cite{amundsen14,amundsen17,drummond16,goyal18,goyal20,tremblin15,tremblin16,tremblin17cloudless}.

Throughout this paper, we use the term ``troposphere'' to refer to the deep convective region of the atmosphere and the term ``stratosphere'' to refer to 
the region above the radiative-convective boundary, where radiative processes dominate the vertical transport of energy.  Although we do not consider a 
``thermosphere'' in our models, the stratosphere would end (and the thermosphere would begin) near the homopause level of the major molecular constituents 
--- at altitudes above that level, atomic species would dominate and atmospheric temperatures would increase significantly.  Moses et al. \cite{moses11} 
demonstrate that the presence of a high-temperature thermosphere does not affect the stratospheric photochemical-model results for H$_2$-rich exoplanets, 
thus justifying our omission of the hot thermosphere.

\subsection{Atmospheric chemistry model} \label{sec:chem}

Thermochemical equilibrium was assumed for the composition of the planets in section \ref{sec:2datmo}.  While that is a reasonable first-order 
assumption, disequilibrium chemical processes such as photochemistry and transport-induced quenching can alter the atmospheric composition of 
H$_2$-rich exoplanets
\cite{agundez14pseudo,agundez12,agundez14gj436,blumenthal18,drummond16,hobbs19,hu14,kawashima18,kawashima19,kitzmann18,koppa12,koskinen13a,koskinen13b,lavvas14alkali,line11gj436,lupu16,miguel14,miguel15,molaverdikhani19,moses13gj436,moses13coratio,moses16,moses11,rimmer16,shulyak20,tsai18relax,tsai17,venot14,venot18ariel,venot13,venot12,zahnle16,zahnle14}.
Such disequilibrium processes are important to include when considering global variations in composition that could affect phase-curve 
observations \cite{agundez14pseudo,drummond20,drummond18,mendonca18,steinrueck19,venot20wasp43}.  
We have therefore converted the 1D Caltech/JPL \texttt{KINETICS} model \cite{allen81,yung84,moses11} to a pseudo-2D chemical model, following the procedures of Ag{\'u}ndez et al. 
\cite{agundez14pseudo}, to track the longitude-altitude variation in composition on our grid of exo-Neptunes.   \texttt{KINETICS} was first used in this 
pseudo-2D mode in Venot et al. \cite{venot20wasp43}.   

In pseudo-2D, \texttt{KINETICS} solves the one-dimensional (vertical) continuity equations as a function of time as the atmospheric column 
``rotates'' around the planet within the low-latitude zonal wind jet, with the physical background conditions changing at each longitude due to the 
temperature-structure and irradiation-angle variations as a function of longitude.
The vertical pressure grid remains constant with longitude in the model, but the densities and altitude scaling between the grid points vary 
as the atmosphere expands or contracts due to the higher or lower temperatures.  We use equation (48) of Showman et al. \cite{showman10} to estimate 
the zonal jet speeds on these planets, which in turn define the pseudo-rotation rates and the timing of the gas residence at each longitude grid point.  The 
rotation rate is assumed to be constant with altitude.  Based on the discussion in Ag{\'u}ndez et al. \cite{agundez14pseudo}, we start the models with 
a fully converged 1D thermo/photochemical model run at the longitude of the hottest stratospheric temperatures on the planet (ranging from 54\deg\ eastward 
of the substellar longitude for the coolest planet to 0\deg\ for the hottest planet, but that hot-spot offset longitude also depends on pressure).  The 
time that the model remains at the temperature-pressure conditions for any particular longitude is based on the pseudo-rotation rate and the longitude 
grid spacing, with the stellar zenith angle and temperature structure remaining fixed for that amount of time.  The species mixing ratio profiles from 
the end of the run at one longitude are then passed as initial conditions for the run at the next longitude.  The calculations continue for multiple 
planetary years, until each year produces repeatable results --- an overall amount of time that is typically controlled by how quickly the photochemical 
species produced in the upper atmosphere are transported to deeper, hotter regions of the atmosphere, where all species are converted back to 
equilibrium. 

The temperature fields from the \texttt{2D-ATMO} calculations described in section \ref{sec:2datmo} are used to define the background atmospheric structure 
for the chemical models, using the same 18\deg\ longitude grid spacing.  However, the vertical grid in the chemical model is extended to lower pressures 
(to our top boundary at 10$^{-8}$ mbar) to reliably capture the effects of high-altitude, ultraviolet-driven photochemistry.  This high-altitude extension is 
accomplished by assuming a simple (and arbitrary) power law extrapolation from the top pressures in the \texttt{2D-ATMO} results.  Occasionally, the profiles also 
require extension to deeper pressure levels to capture the N$_2$-NH$_3$ quench point, in which case we assume a straight-line extrapolation in temperature 
with $\log P$.  If the \texttt{2D-ATMO} temperatures exceed $\sim$2900-3000 K in the deep atmosphere, we truncate the profiles at those temperatures to avoid 
numerical instability in the chemical calculations.  These extrapolated and/or truncated temperature profiles are then run through a hydrostatic-equilibrium 
routine to set up a background atmospheric grid of $\sim$200 vertical levels for each longitude across each planet, using the same planetary parameters described 
in section \ref{sec:2datmo}.  The temperatures along this pre-calculated longitude-altitude grid are then held fixed throughout the chemical calculations; 
that is, any composition changes due to the disequilibrium chemistry do not feed back to affect the temperatures (see \cite{drummond16,hu14} for evaluations 
of the reasonableness of this assumption).  

A large complement of 277 atmospheric species composed of 23 different elements has been considered in the thermochemical-equilibrium modeling used to 
define the 2D thermal structure for our planetary grid in section \ref{sec:2datmo}.  However, kinetics data (i.e., chemical reaction rates, 
product pathways and branching ratios) for many of these elements and species are lacking, causing us to restrict our chemical-kinetics calculations 
to a smaller number of species/elements.  Here, we consider the 
kinetics of C-, O-, N-, and H-bearing species.  We adopt the chemical network of Moses et al. \cite{moses13gj436}, in which H, H$_2$, He, and 90 other 
species with up to six carbon atoms, three oxygen atoms, and two nitrogen atoms interact via $\sim$1600 kinetic reactions ($\sim$800 fully reversed 
reaction pairs).  The full reversal of the reaction mechanism allows thermochemical equilibrium to be reproduced when chemical time constants are 
shorter than the transport time scales \cite{moses11}.  Photolysis reactions are included, but are not reversed, as photodissociation is inherently a disequilibrium 
process.  The stellar ultraviolet flux adopted for the calculations is described in detail in Venot et al. \cite{venot20wasp43}.  A solar composition is 
assumed for the planetary atmospheres in our nominal models \cite{caffau11}, but we also test the sensitivity of the results to higher atmospheric 
metallicities. Condensation is 
not considered in the kinetics models presented here, as none of our C-, N-, and O-bearing species are expected to condense, but for cases in which 
the thermochemical-equilibrium calculations determine that silicates and metal oxides will condense at depths below the infrared ``photosphere'' of 
our planets, we reduce the oxygen abundance by an appropriate amount (determined by the thermochemical-equilibrium calculations with a full complement 
of species and elements) to account for the fraction of the oxygen being sequestered in refractory condensates.  This fraction varies with metallicity, 
but is of order 20\%.  For cases in which the temperatures derived from \texttt{2D-ATMO} at all longitudes are found to reside above the condensation curve for 
the major magnesium-silicate condensates, the oxygen abundance is not reduced.

\begin{figure*}[!htb]
\begin{center}
\includegraphics[width=0.9\textwidth]{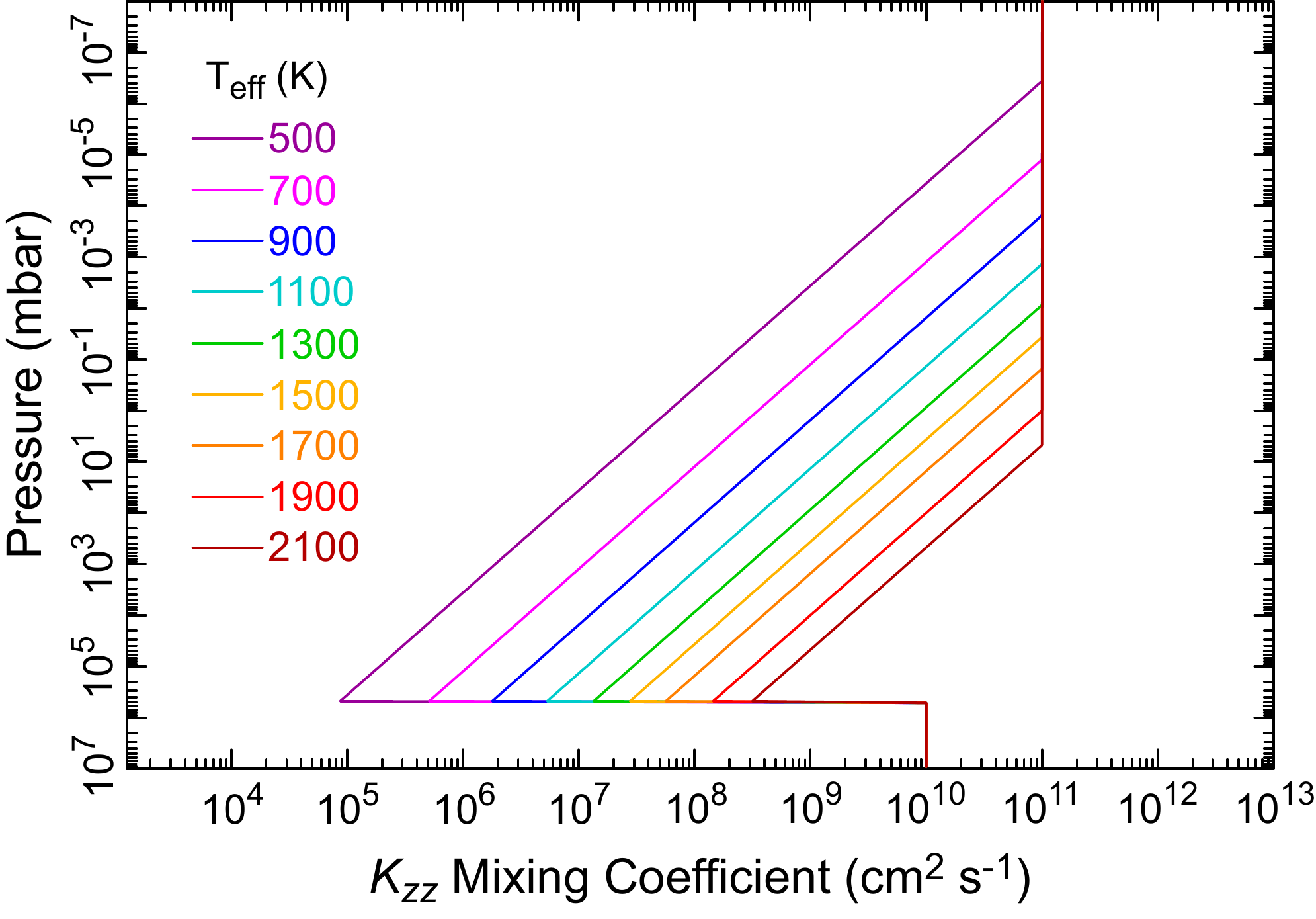}
\end{center}
\vspace{-0.5cm}
\caption{$K_{zz}$ diffusion coefficient profiles adopted for our grid of exo-Neptunes exoplanets with different \teff .}
\label{figeddyall}       
\end{figure*}

\begin{figure*}[!htb]
\begin{center}
\includegraphics[width=0.85\textwidth]{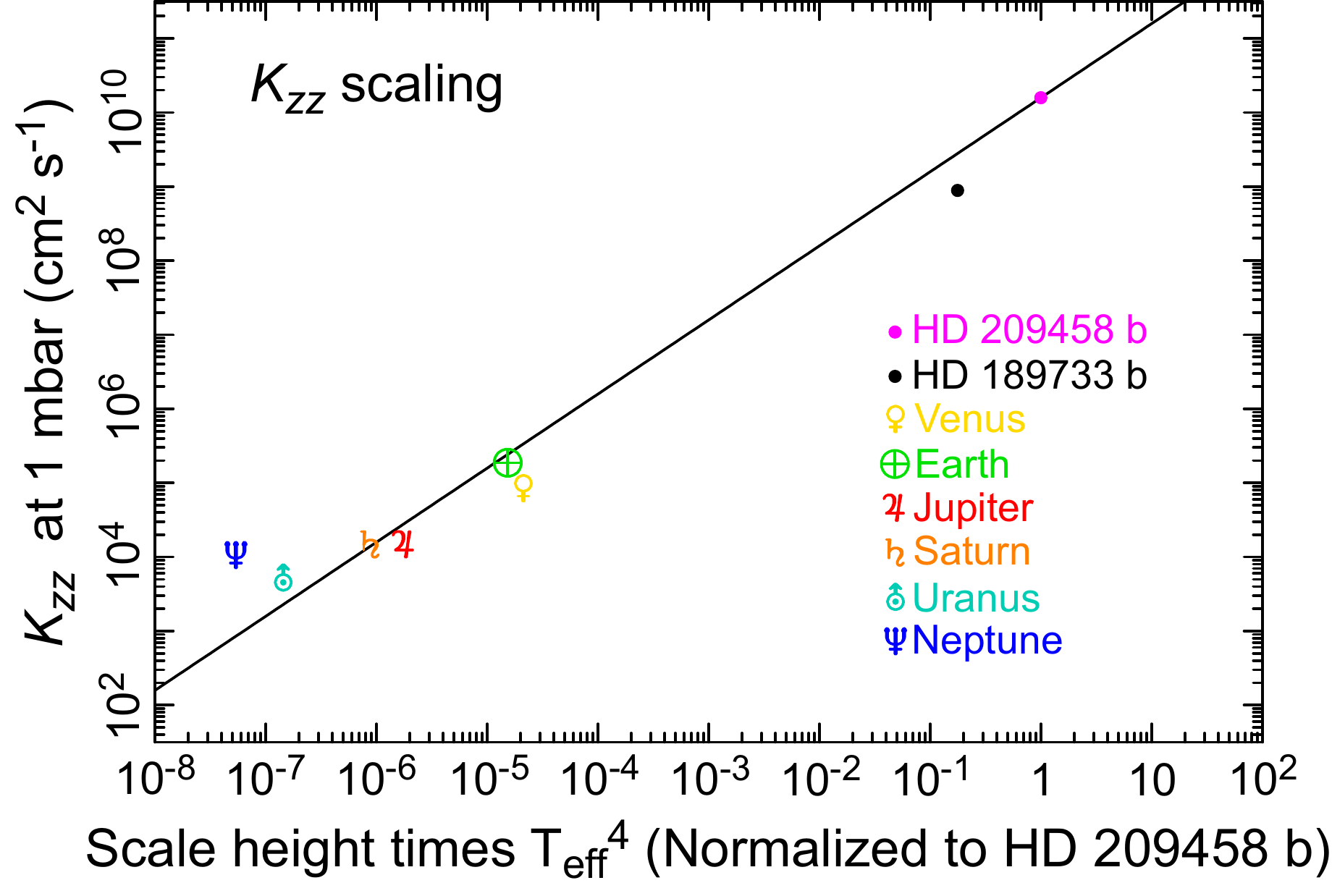}
\end{center}
\vspace{-0.5cm}
\caption{The 1-mbar $K_{zz}$ diffusion coefficient derived from Eq.~(1) (solid black line) is compared with values derived from 3D-GCM tracer studies 
of HD 209458 b \cite{parmentier13} and HD 189733 b \cite{agundez14pseudo}, and compared with values derived from 1D 
global-average photochemical models of solar-system planets \cite{moses05,yung99,zhangshow18a} (colored points, as labeled).  For the solar-system 
planets, a global-average $H_\mathrm{{1mbar}}$ is used to calculate the scaling factor along the abscissa, and the scaling factors have been normalized 
such that the value for HD 209458 b is unity.}
\label{figeddyscale}       
\end{figure*}

Vertical transport in 
the model occurs via eddy and molecular diffusion.  The molecular diffusion coefficients are taken from Moses et al. \cite{moses00a}.  The eddy diffusion 
coefficient ($K_{zz}$) profiles for the different planets are shown in Fig.~\ref{figeddyall}.  In the deep, convective region of the atmosphere, we adopt 
a constant, large $K_{zz}$ value of 10$^{10}$ cm$^{2}$ s$^{-1}$ for all the planets, based on free-convection and mixing-length theories 
\cite{visscher10co}.  At high altitudes, we restrict $K_{zz}$ from exceeding 10$^{11}$ cm$^{2}$ s$^{-1}$, based on limits observed in 3D models 
and actual atmospheres \cite{freytag10,parmentier13,yung99}.  In the intermediate radiatively-controlled stratospheric region, atmospheric mixing is 
typically dominated by atmospheric waves, which results in $K_{zz}$ varying with the inverse square root of atmospheric pressure \cite{lindzen81}.  
The magnitude of that mixing, however, depends on energy from the planet's interior (which we assume is small compared to the external energy 
source and is the same for all the planets), as well as the energy received from the absorption of stellar radiation (which varies with orbital distance 
and thus planetary \teff ) \cite{freytag10}.  
Based on 3D GCM tracer transport and the discussions and analytic expressions in Freytag et al. \cite{freytag10}, Parmentier et al. \cite{parmentier13}, 
Zhang \& Showman \cite{zhangshow18b}, and Komacek et al. \cite{komacek19}, we assume that the stratospheric $K_{zz}$ scales as 
\begin{equation}
K_{zz} \ = \ 5\, \times 10^8 \, \left[ P\hbox{(bar)} \right]^{-0.5} \, \left( \frac{H_\mathrm{{1mbar}}}{620 \hbox{ km}} \right) 
\left( \frac{T_\mathrm{{eff}}}{1450 \hbox{ K}} \right)^4 \quad ,
\end{equation}
where K$_{zz}$ is in units of cm$^2$ s$^{-1}$, $P$ is the atmospheric pressure (in bar), and $H_\mathrm{{1mbar}}$ is the atmospheric pressure scale height 
at 1 mbar (in km), taken from the evening terminator of our \texttt{2D-ATMO} models.  This expression follows the GCM-derived vertical profile of $K_{zz}$ for 
HD 209458 b from studies of tracer transport \cite{parmentier13}, and the scaling to other planetary \teff\ provides a reasonable fit to the 
stratospheric eddy diffusion coefficients 
inferred for solar-system planets from 1D photochemical models \cite{moses05,yung99,zhangshow18a} (see Fig.~\ref{figeddyscale}).  
For boundary conditions, we assume zero flux at both the top and bottom boundary.

We also calculate thermochemical-equilibrium solutions as a function of altitude and longitude for our grid of planets using the same temperature fields,  
the same (reduced) set of chemical species and elements used in the pseudo-2D kinetics model, and the same assumptions about bulk elemental abundances, so that 
we can directly compare the results for equilibrium versus disequilibrium chemistry on these planets.  The CEA code from Gordon \& McBride \cite{gordon94} 
is used to calculate thermochemical equilibrium, in this case.

\subsection{Planetary emission model} \label{sec:emission}

Once we obtain the longitude- and altitude-dependent abundance profiles for the atmospheric constituents from the chemical models described in 
section \ref{sec:chem}, we reconstruct a global composition assuming that the mixing ratios derived for the equatorial region in our 
pseudo-2D models are latitude invariant.  For this reconstruction we also add back in chemical-equilibrium abundances for all species such 
as Na, K, TiO, H$_2$S, etc. (see section \ref{sec:2datmo}) that were not considered in the pseudo-2D kinetics model.  Based on 3D GCMs for 
hot Jupiters that consider simplified chemistry \cite{mendonca18,drummond20}, the assumption of latitude invariance appears to be justified for 
cooler exoplanets with equilibrium temperatures $\lta$ 1100 K; however, latitude variations in species' abundances --- particularly between the equatorial jet 
region and the rest of the planet --- do tend to be significant for hotter planets with equilibrium temperatures $\gta$ 1300 K.  As we show from our model 
results below, thermochemical equilibrium tends to dominate the composition and phase-curve behavior of these hotter planets, in any case, so 
that pseudo-2D models are not required to accurately predict the spectral behavior of such planets.  Based on the reconstructed global 
composition, we then use \texttt{ATMO} to calculate the disk-averaged planetary emission spectrum at each orbital phase by 
summing up the outgoing intensity from each of the 18\deg\ longitude by 30\deg\ latitude geographical regions that are located on the observable hemisphere 
of the planet at each position in the planet's orbit.  We perform these calculations for wavelengths from 0.5-11.8 $\mu$m, assuming a constant wavenumber 
resolution of 100 cm$^{-1}$.  Note that this wavelength range goes beyond the maximum wavelength explored by \textit{Ariel} (7.8 $\mu$m), so that we can 
highlight some interesting phase-curve behavior at longer wavelengths.

\section{Results} \label{sec:results}

\subsection{2D thermal structure results} \label{sec:tempresults}

The results from the \texttt{2D-ATMO} thermal-structure calculations are shown in Fig.~\ref{figtemp}.  Note that day-side atmospheric temperatures and day-night 
temperature differences in our models become larger for planets at smaller orbital distances (i.e., larger planetary \teff ).  Thermal inversions often 
occur at day-side longitudes, and become particularly strong at larger planetary \teff .  Such inversions are caused by the absorption of short-wavelength 
stellar radiation (especially in the visible) by atmospheric constituents such as TiO, VO, Fe, and alkalis for the warmer planets where these species 
have not condensed at depth \cite{hubeny03,fort08uni,goyal20}, by H$^{-}$ for hot planets \cite{arcangeli18,goyal20}, and by the short radiative time 
constants at low pressures that cause the atmosphere to respond relatively rapidly to the strong dayside instellation variations \cite{iro10}.  
For all \teff , the hottest regions of the atmosphere can be found either right at (for the hottest planets) or downwind (i.e., eastward, for the cooler 
planets) of the substellar longitude, although the actual longitude of the temperature 
maxima can change with pressure due to radiative cooling rates that vary with pressure.  The dashed curve in Fig.~\ref{figtemp} illustrates where CO and 
CH$_4$ would have equal abundance in thermochemical equilibrium for a solar composition atmosphere.  Note that the temperature profiles at all longitudes 
for the \teff\ = 500 K planet reside completely in the CH$_4$-dominated regime (i.e., CH$_4$ would be the dominant equilibrium carbon 
constituent throughout the atmosphere), whereas the temperature profiles for the planets with \teff\ $\ge$ 1700 K are located completely 
in the CO-dominated regime (i.e., CO would be the dominant equilibrium carbon constituent everywhere).  At 
intermediate \teff , the temperature profiles cross the CO-CH$_4$ equal-abundance curve at least once, suggesting that if thermochemical 
equilibrium were to prevail, the dominant carbon-bearing component in the atmosphere would switch from CH$_4$ to CO within some region(s) of the
atmosphere.  For 700 K $\le$ \teff\ $\le$ 1100 K, CH$_4$ dominates in some portion of the stratosphere during the nighttime, while CO dominates 
during the daytime.  

\begin{figure*}[!htb]
\begin{center}
\includegraphics[width=0.49\textwidth]{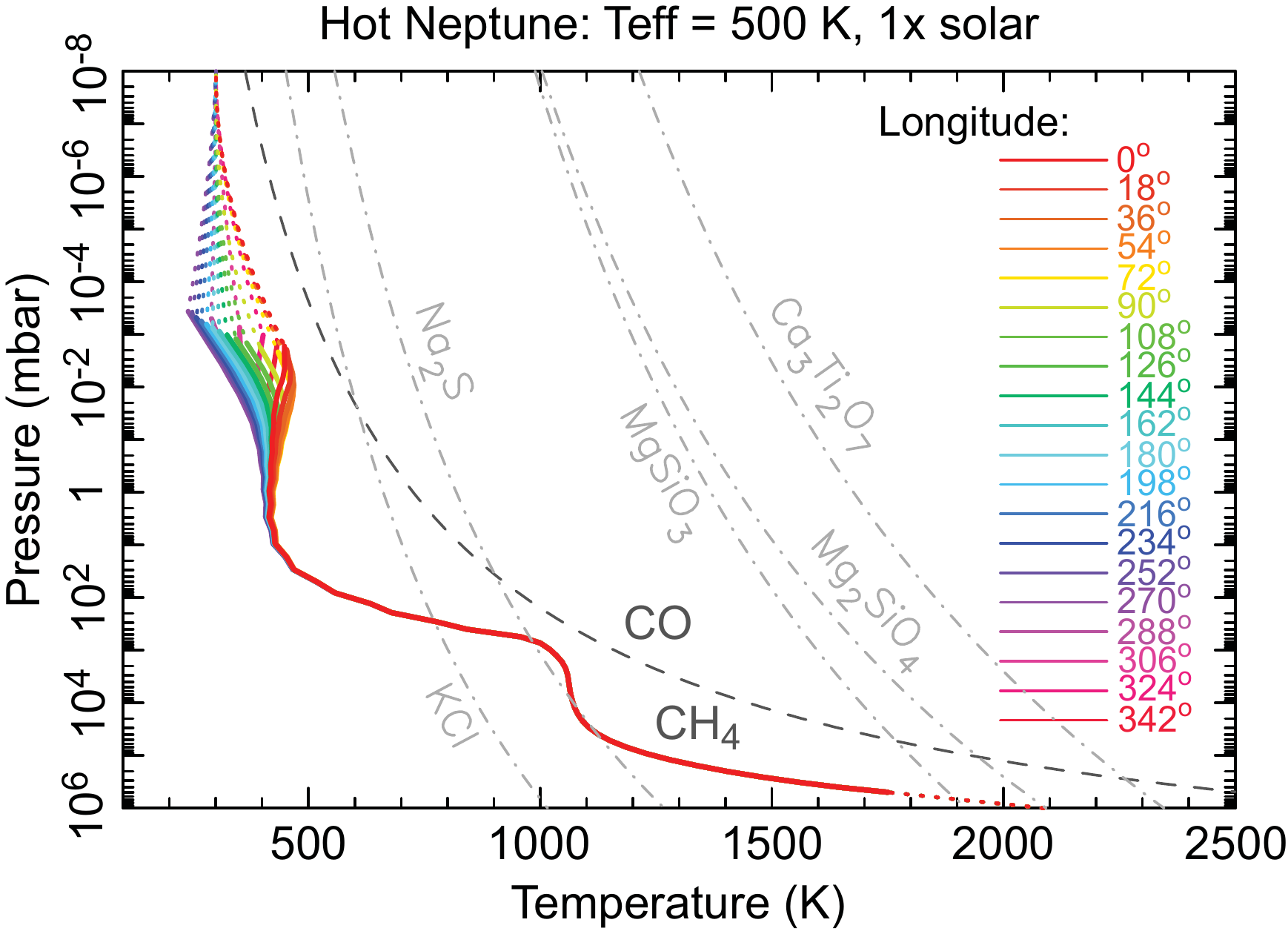} 
\includegraphics[width=0.49\textwidth]{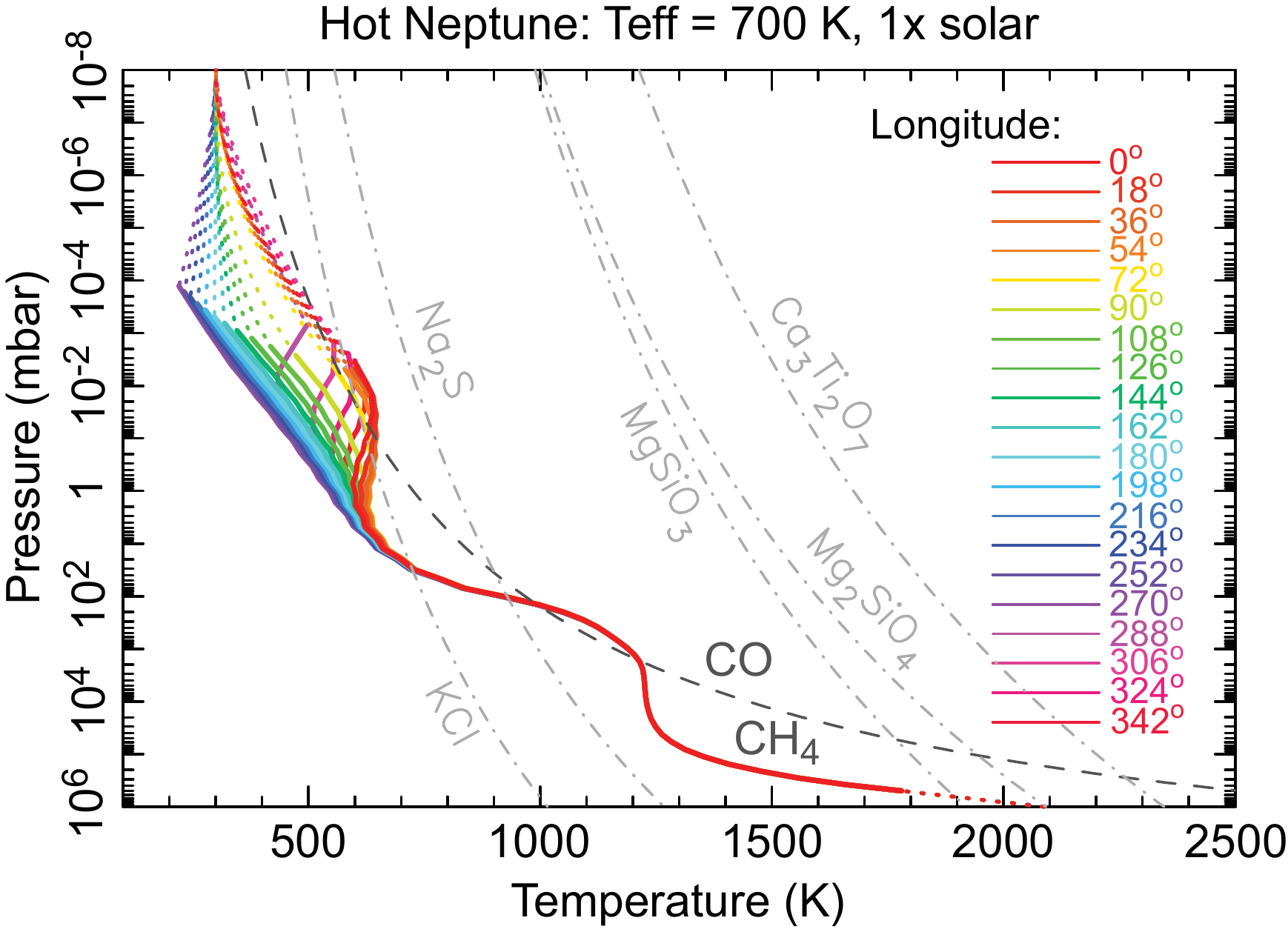} \\
\vspace{10pt}
\includegraphics[width=0.49\textwidth]{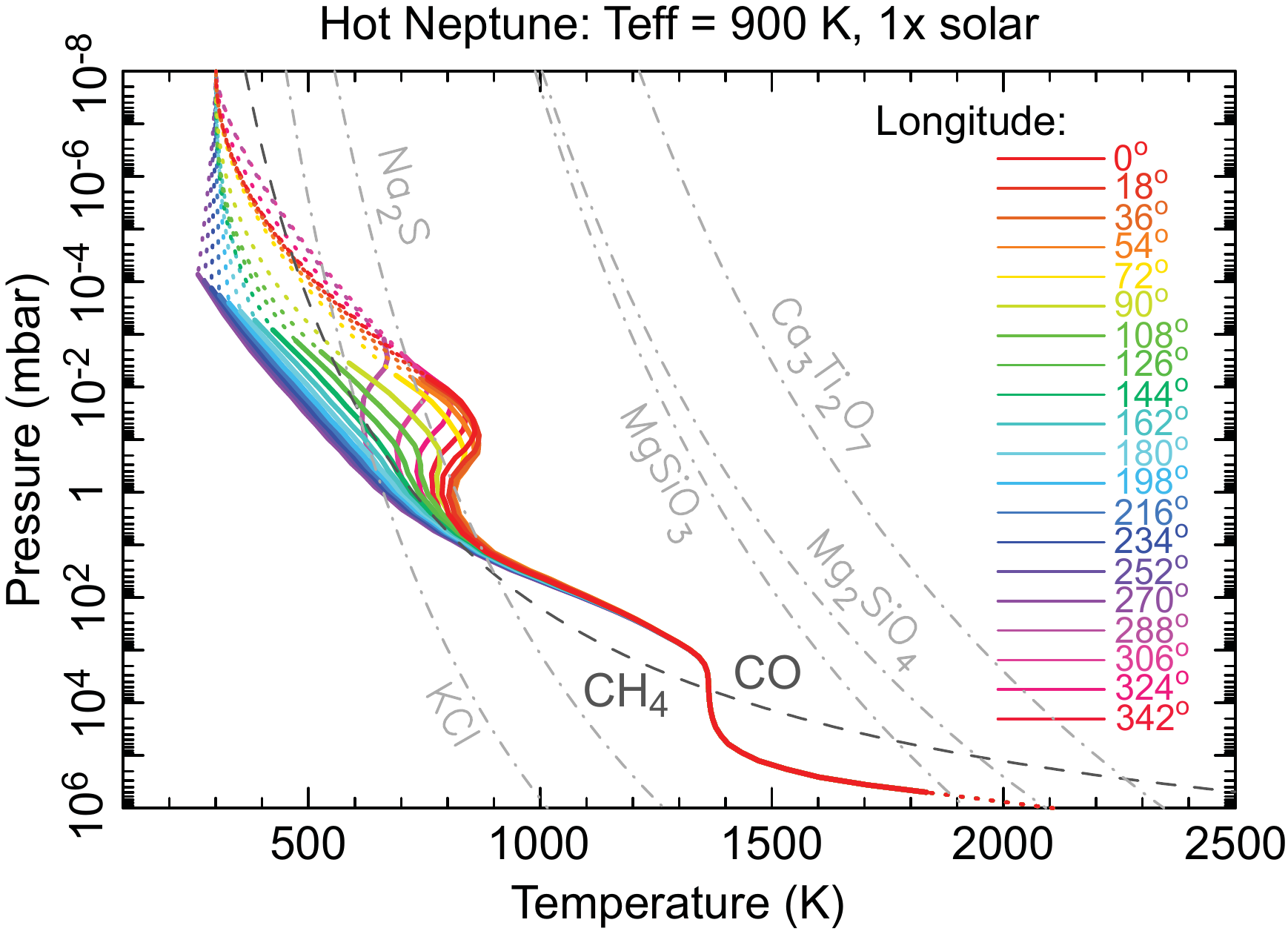} 
\includegraphics[width=0.49\textwidth]{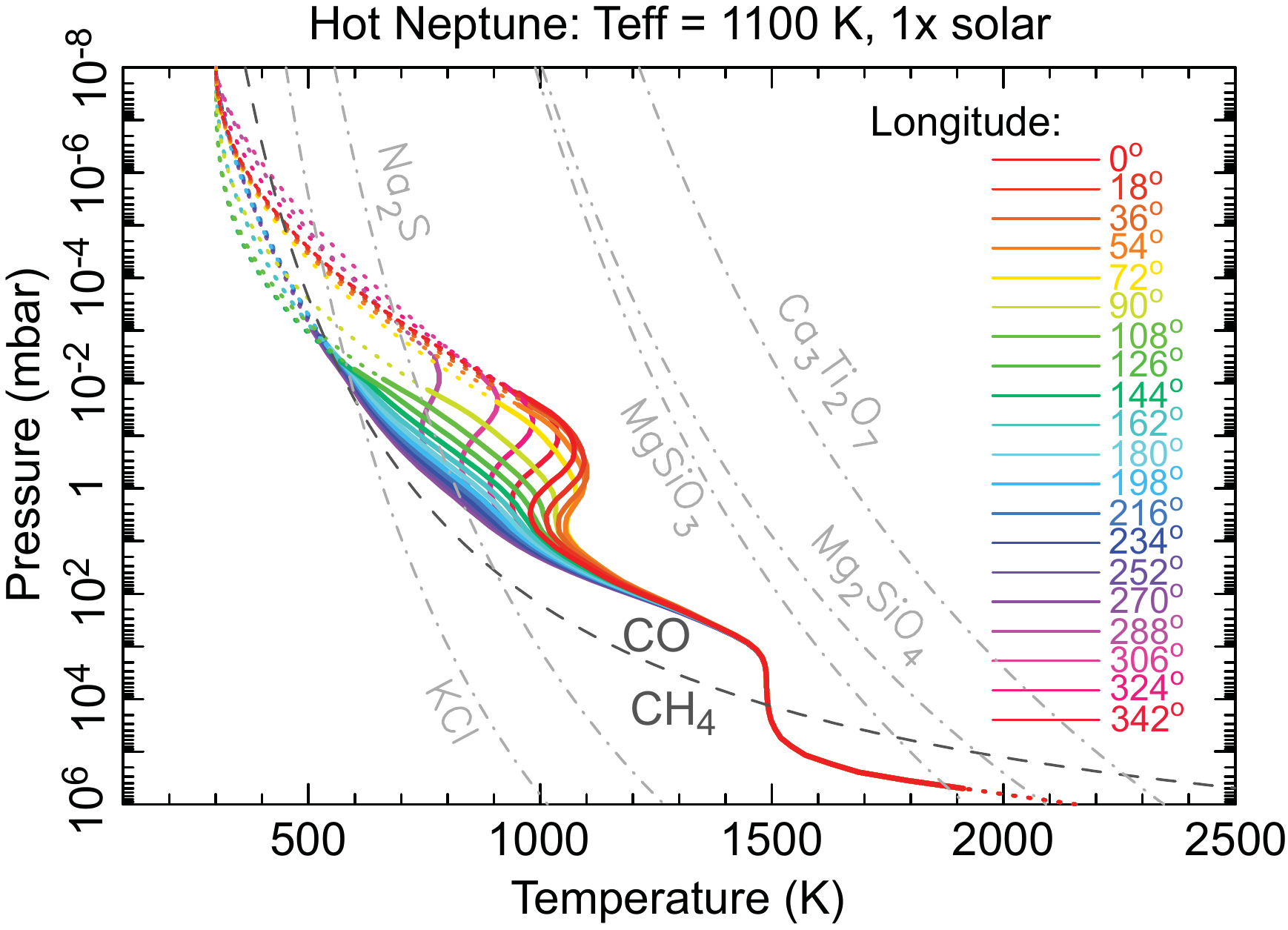} \\
\vspace{10pt}
\includegraphics[width=0.49\textwidth]{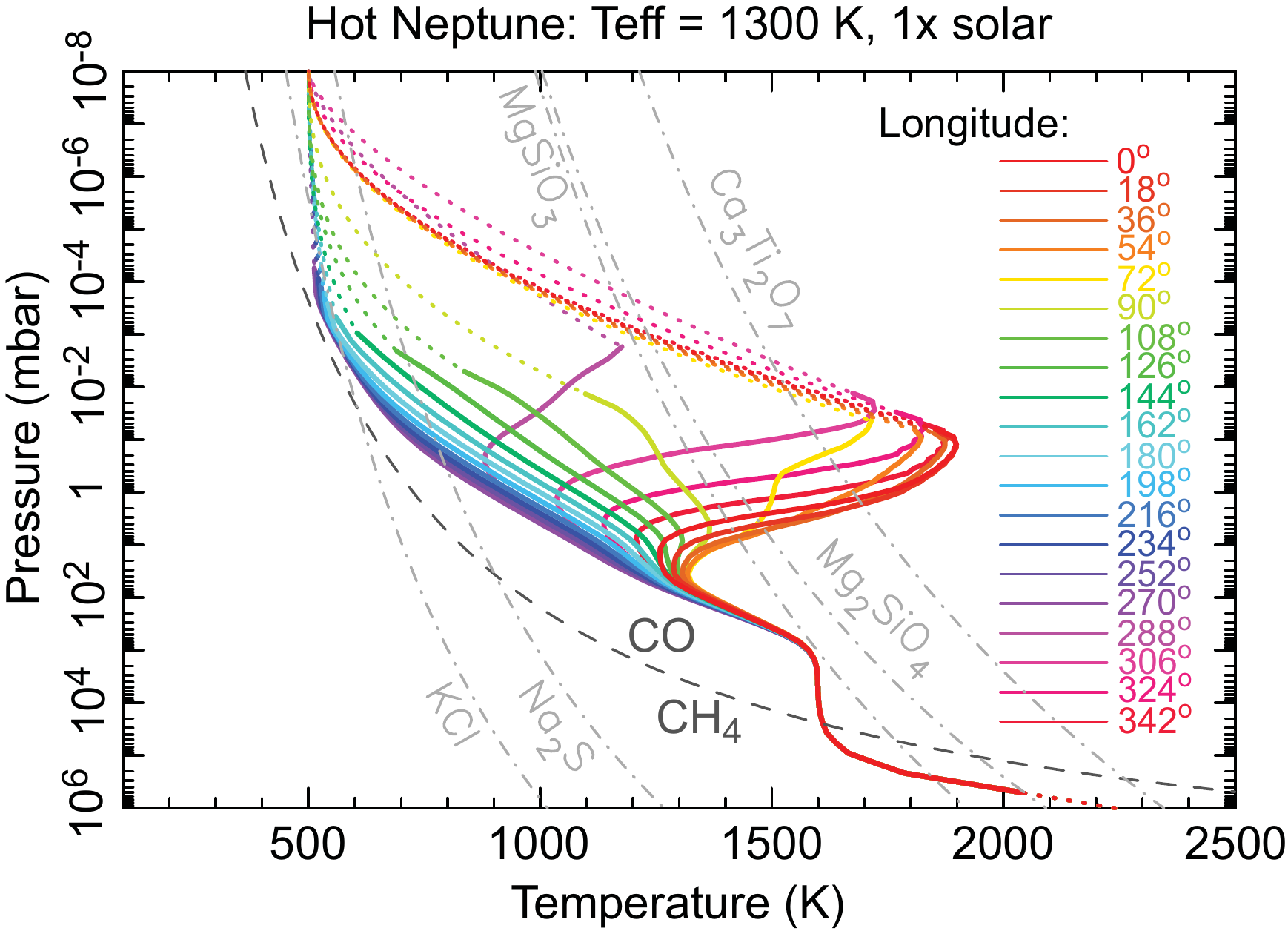}
\includegraphics[width=0.49\textwidth]{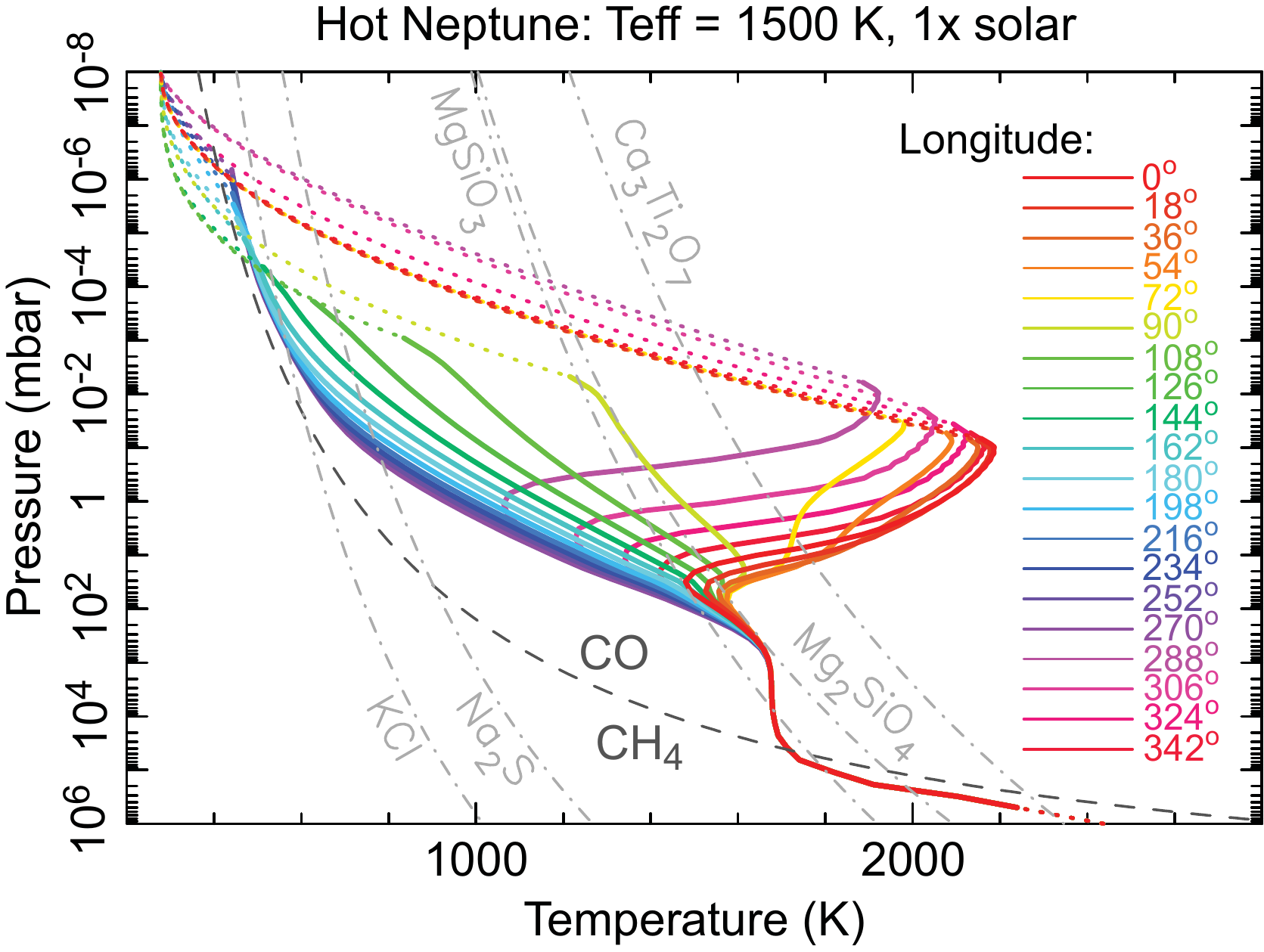} 
\end{center}
\vspace{-0.5cm}
\noindent{Fig.~\ref{figtemp}. \textit{(continued)}.}
\end{figure*}

\clearpage

\begin{figure*}[!htb]
\begin{center}
\includegraphics[width=0.49\textwidth]{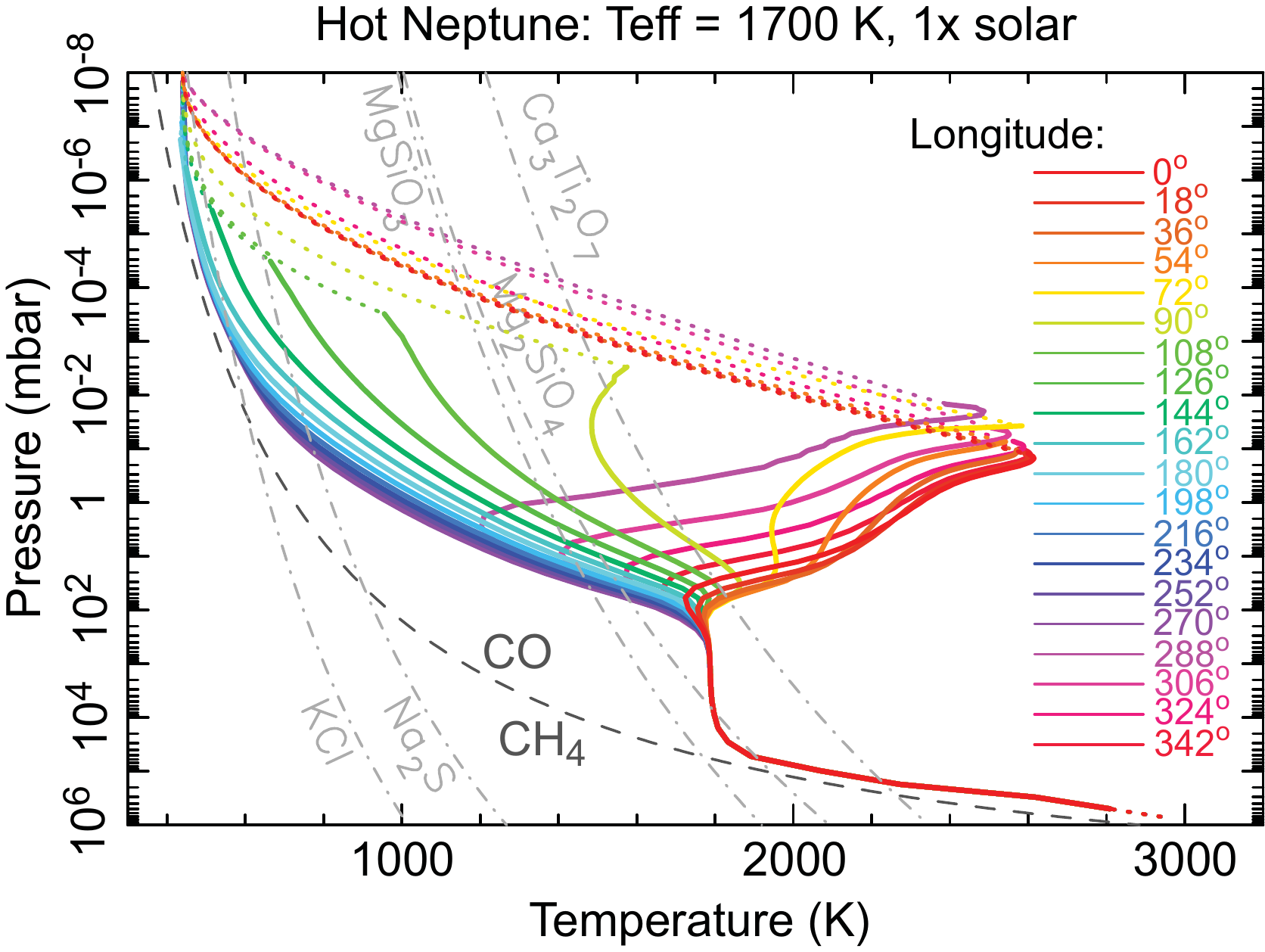}
\includegraphics[width=0.49\textwidth]{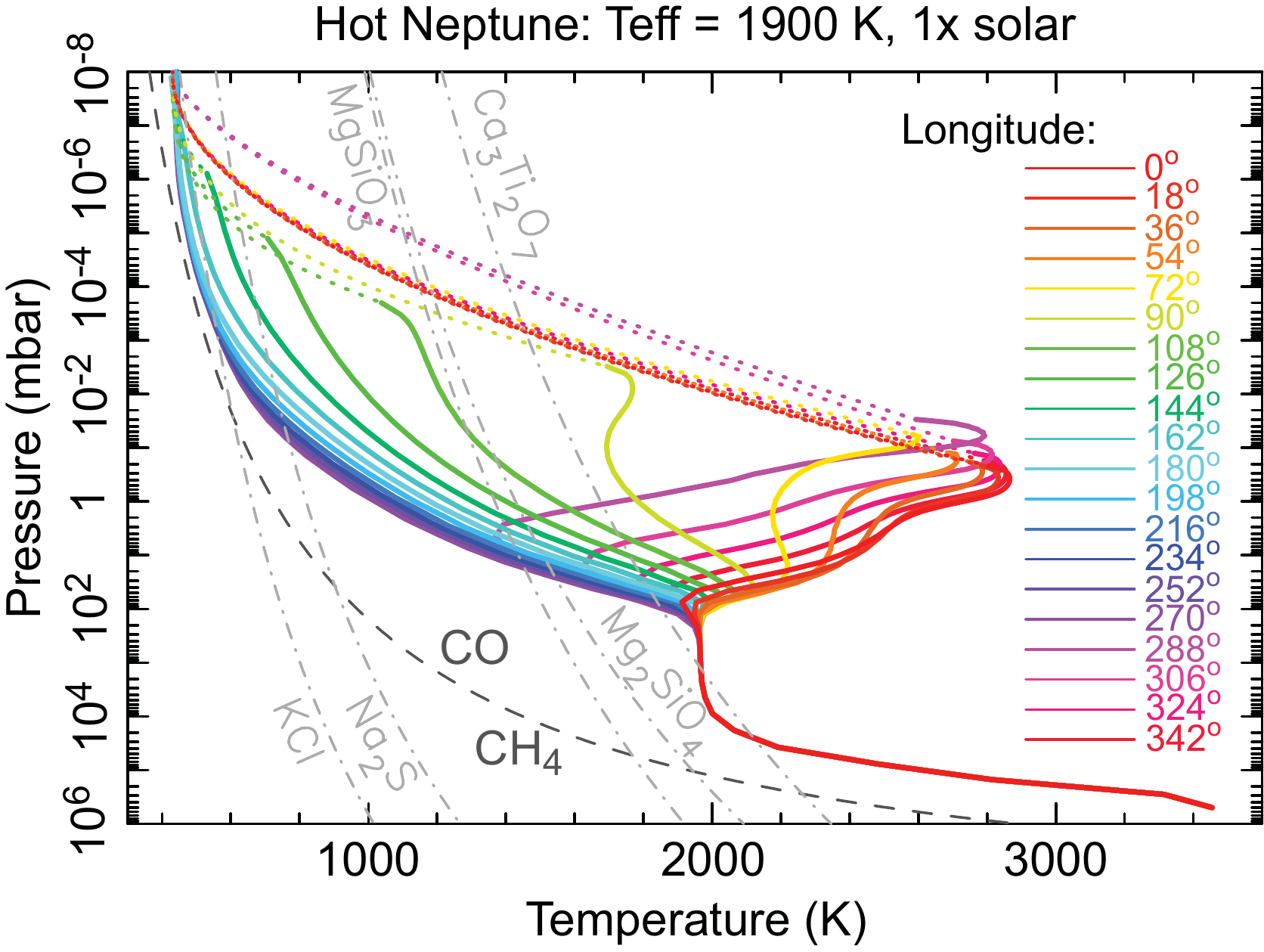} \\
\vspace{10pt}
\includegraphics[width=0.49\textwidth]{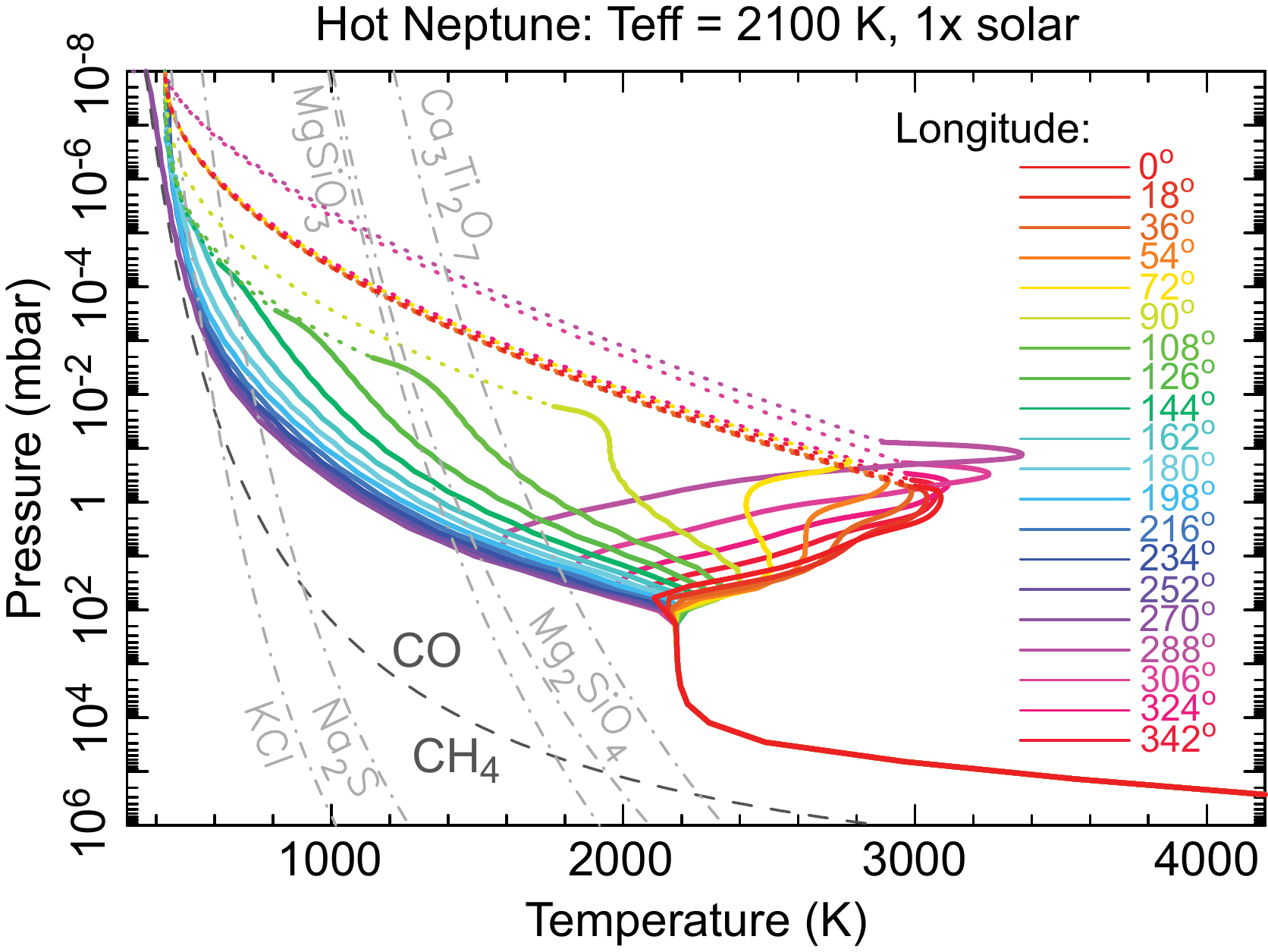}
\end{center}
\vspace{-0.5cm}
\caption{Equatorial temperature-pressure profiles as a function of longitude (colored solid lines, as labeled; colored dotted lines indicate extrapolations 
beyond the original \texttt{2D-ATMO} calculations) for the nine exo-Neptunes in 
our grid, assuming a solar composition.  The planetary effective temperature \teff\ is listed at the top of each image.  The substellar point on the 
day-side hemisphere is at longitude = 0$^{\circ}$.  Also shown in the figures 
are the condensation curves for some major equilibrium cloud condensates (gray dot-dashed lines) and the equal-abundance curve for gas-phase CO and 
CH$_4$ (dashed black line). In thermochemical equilibrium, CH$_4$ is the dominant carbon constituent in regions where the temperature profile extends 
below that curve, whereas CO becomes dominant when the temperature profile extends above the curve.}
\label{figtemp}       
\end{figure*}

\begin{figure*}[!htb]
\begin{center}
\includegraphics[width=0.48\textwidth]{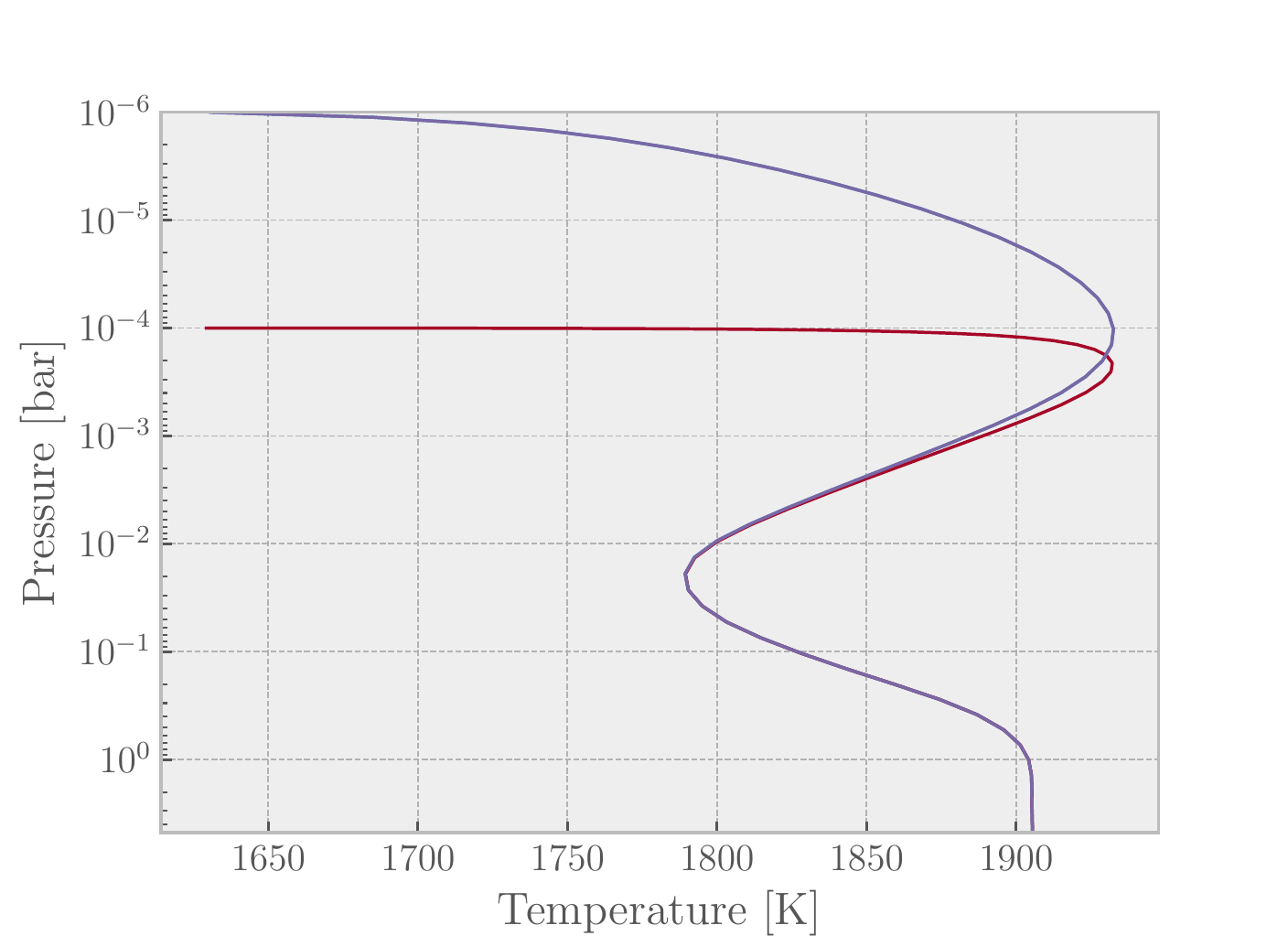}
\includegraphics[width=0.50\textwidth]{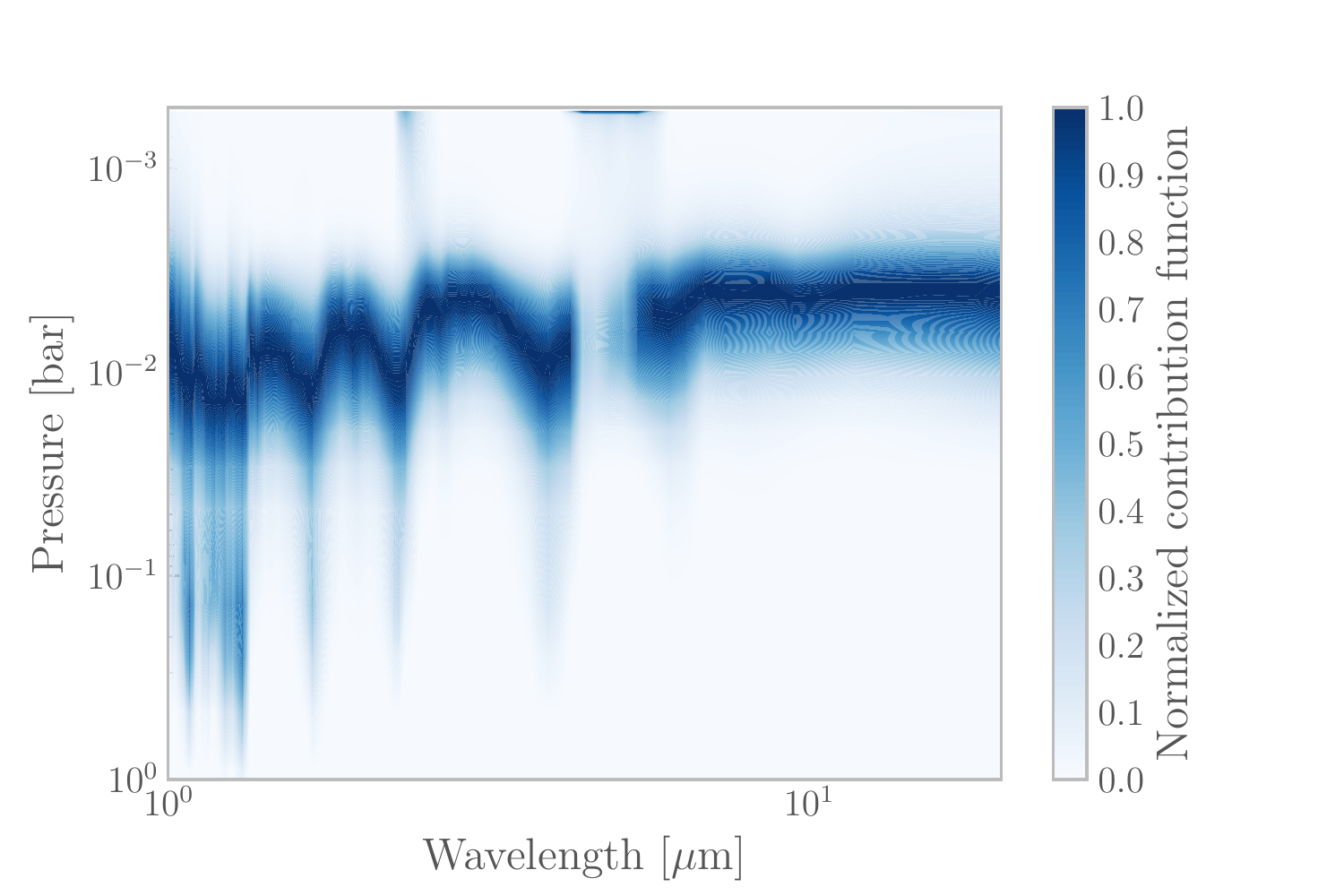} 
\end{center}
\caption{Sensitivity of thermal structure to placement of top boundary: (Left) Temperature-pressure profiles calculated for our \teff\ = 2100 K solar-metallicity 
exo-Neptune assuming a top boundary at 10$^{-4}$ bar (red curve) and 10$^{-6}$ bar (blue curve); (Right) Contribution functions for the infrared emission in the 
\teff\ = 2100 K model with the shallower boundary (i.e., red model at left).}
\label{figtemptest}       
\end{figure*}

Because the \texttt{2D-ATMO} models were calculated with altitude as the vertical coordinate, the pressure at the top of the atmosphere in these models 
varies with longitude.  The top pressure is particularly large for the dayside atmospheres of the hotter planets, raising some concerns that the top 
pressure cut-off could be introducing boundary-condition sensitivities that could be affecting the derived pressure-temperature profiles or even 
the magnitude of the temperature inversion, especially for the high \teff\ cases.  To investigate this possibility, we performed 1D radiative-transfer 
calculations for the \teff\ = 2100 K case (using \texttt{ATMO}, see \cite{goyal20}) to determine how sensitive the pressure-temperature profile is 
to the assumed upper boundary of the model.  Thermochemical equilibrium is assumed for these tests.  The results are shown in Fig.~\ref{figtemptest}, 
along with the emission contribution functions for the shallower model with the higher-pressure cut-off.  Because most of the energy is absorbed and 
re-emitted in the $\sim$3$\scinot-3.$ bar region or deeper, we find only minor changes in the predicted temperatures at pressures greater than 
$\sim$10$^{-3}$ bar.  Temperature differences do become more significant near the top boundary of the shallower model, but the overall magnitude of 
the temperature inversion is not significantly affected.  Fig.~\ref{figtemptest} demonstrates that peaks in the contribution function do occur at 
the top of the model at wavelengths of $\sim$2.3--2.5 and $\sim$4.5--5 $\mu$m due to optically thick CO bands in those regions (which incidentally 
also highlight the important role that CO plays in regulating stratospheric temperatures on hot Neptunes).  However, the 
placement of the top boundary mainly influences the pressure at which this CO emission originates, not the temperature; because pressure 
broadening is weak at these pressures, the impact of the shallower boundary on the predicted emission is relatively minor.  

\subsection{Chemical-modeling results} \label{sec:chemresults}

The results regarding the changes in atmospheric composition as a function of pressure and longitude are shown in Fig.~\ref{figmixall}.  
Here we show both the thermochemical-equilibrium results and the pseudo-2D chemical model results side by side.  The two solutions are similar 
in the lower atmosphere, where temperatures are high enough that thermochemical equilibrium can be maintained kinetically.  Strong differences between 
the equilibrium and pseudo-2D cases begin to appear in the middle atmosphere.  Chemical conversion between different molecular or atomic forms of an 
element is not instantaneous.  At high temperatures, chemical reactions tend to be fast, because any energy barriers to the reaction can be more easily 
surmounted.  Reactions then tend to occur equally rapidly in both the forward and reverse directions, allowing the kinetics to maintain thermochemical 
equilibrium.  When atmospheric transport is present, however, the constituents can be carried to different atmospheric regions, where temperatures 
and reaction rates differ.  If this transport occurs faster than the chemical reactions can maintain equilibrium, then the composition can be ``quenched'' 
at the point at which the transport time scale becomes shorter than the chemical-kinetics conversion time scale between dominant forms of an element
\cite{moses11,line11gj436,venot12}.
Vertical quenching of CH$_4$, CO, and H$_2$O in the convective region of giant planets and brown dwarfs is one well-known example of such a 
transport-induced quenching process \cite{prinn77,fegley96,saumon00,visscher11,cavalie17}, where the constituent that is not dominant at cooler, 
higher altitudes can become present at greater-than-expected mixing ratios as a result of quenching at depth and then transport up to the cooler observable 
regions.  Quenching in the horizontal direction can also occur when horizontal winds transport species faster than chemical reactions can maintain 
equilibrium \cite{coop06,agundez14pseudo,agundez12,bordwell18,drummond18,drummond20,mendonca18}.  

In our pseudo-2D models, both vertical and horizontal quenching are affecting the predicted abundances.  When the vertical quench point is deeper than 
the deepest pressure at which the strong stellar irradiation causes significant temperature variations with longitude (i.e., $\sim$10--200 mbar in our 
models, depending on planetary \teff ; see Fig.~\ref{figtemp}), then vertical quenching at the hottest point on the day side controls where the species 
mixing ratios first depart from the equilibrium profiles, and the strong zonal winds carry the constituents from the hotter day side to the cooler 
night side more rapidly than the gas can equilibrate, leading to longitudinal homogenization of the species profiles at abundances relevant to the vertical 
quench point on the day side (see also \cite{agundez14pseudo,venot20wasp43}).  Even if the longitudinal transport time scales are shorter than the vertical 
transport time scales at depths below the day-side vertical quench point, the global temperatures at these deeper altitudes are uniform with longitude, 
such that the zonal transport does not drive the abundances away from the longitude-invariant equilibrium abundance profiles.  Vertical quenching is 
then the dominant process controlling the global disequilibrium abundances.  However, the hotter planets can have day-side vertical quench points 
at altitudes above the pressure region where significant longitudinal variations in temperature occur --- which is the case for the 
CO-CH$_4$-H$_2$O quench point for our planets with \teff\ $\ge$ 1500 K, or for the N$_2$-NH$_3$ quench point for our planets with \teff\ $\ge$ 1900 K.
In that situation, the zonal winds can control the departure of the species profiles from equilibrium, and the global abundances can be more complicated 
(see \cite{agundez14pseudo,drummond18,drummond20,mendonca18}); however, the abundance profiles for those models tend to more closely match the equilibrium 
solution everywhere across the planet.  At lower pressures, chemical time scales for photochemical processes can be shorter than the zonal-wind transport 
time scales, so longitudinal differences begin to appear in the pseudo-2D models at high altitudes.

%
%
%
%

For all our exo-Neptune models, the dominant photochemically produced species are atomic H, O, C, N (not shown in Fig.~\ref{figmixall}), as well 
as strongly bonded simple molecules such as HCN, C$_2$H$_2$, and NO.  On cooler planets where CH$_4$ is the major carbon constituent, complex hydrocarbons 
and nitriles such as C$_6$H$_6$ and HC$_3$N are produced in non-trivial amounts.  The photochemically produced species are synthesized mainly in the 
middle-to-upper atmosphere and then are transported downwards, where the higher temperatures eventually cause them to be kinetically converted back 
to the ``parent'' molecules, and upwards to levels where molecular diffusion eventually limits their presence.  The mixing ratios of photochemically 
produced species tend to be sensitive to longitude and track the stellar flux in the pressure region where they are predominantly produced, such that 
the photochemical product abundances in these regions are greater on the day side than the night side.  However, photochemically produced molecules 
can also be destroyed by ultraviolet photolysis, so their abundance at altitudes above their dominant production region often increases on the night-side 
hemisphere.  This high-altitude night-side increase is also aided by the recombination of reactive radicals that are produced at high net steady-state 
rates on the day side, are carried by the winds to the night side, and then react to form new species with their loss rates strongly exceeding their 
production rates.  Photochemical ``parent'' molecules such as H$_2$O, CH$_4$, NH$_3$, N$_2$, and CO are destroyed at mid-to-high altitudes by photolysis 
and by reactions with atomic H and other species produced by photochemistry, so their high-altitude mixing ratios are reduced on the day side and recover 
on the night side.  

\begin{figure*}[!htb]
\begin{center}
\includegraphics[width=1.0\textwidth]{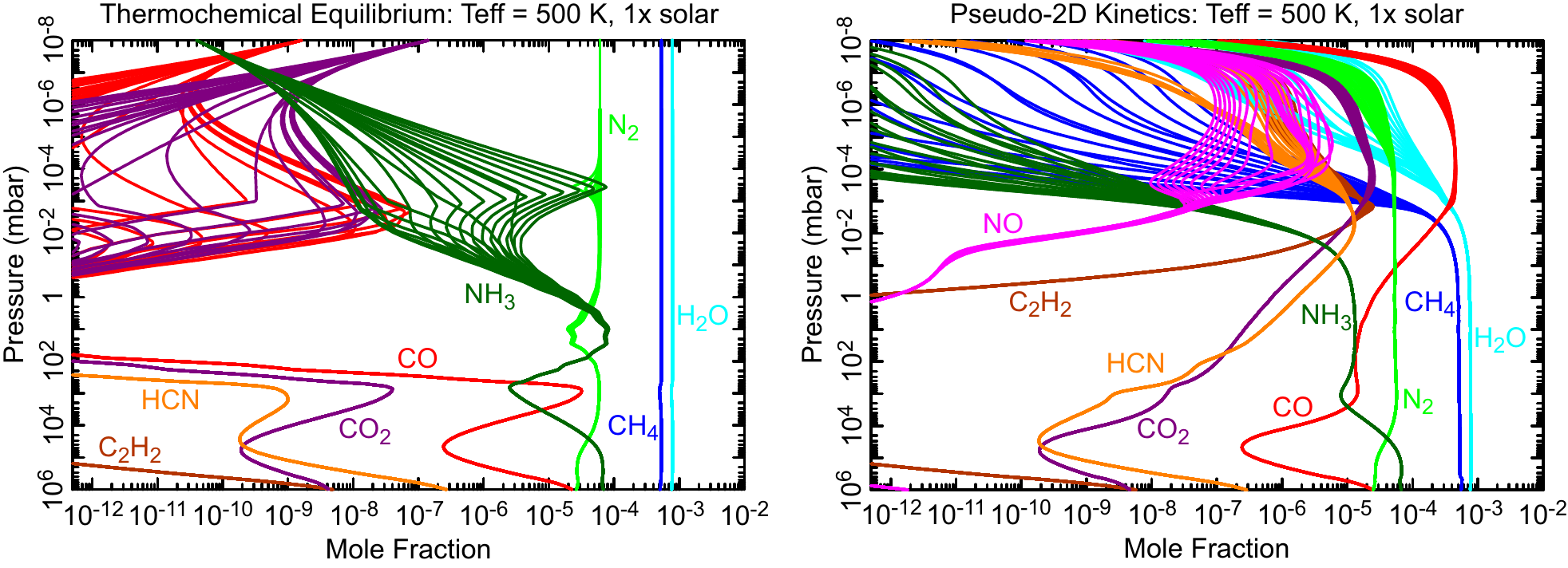} \\
\vspace{10pt}
\includegraphics[width=1.0\textwidth]{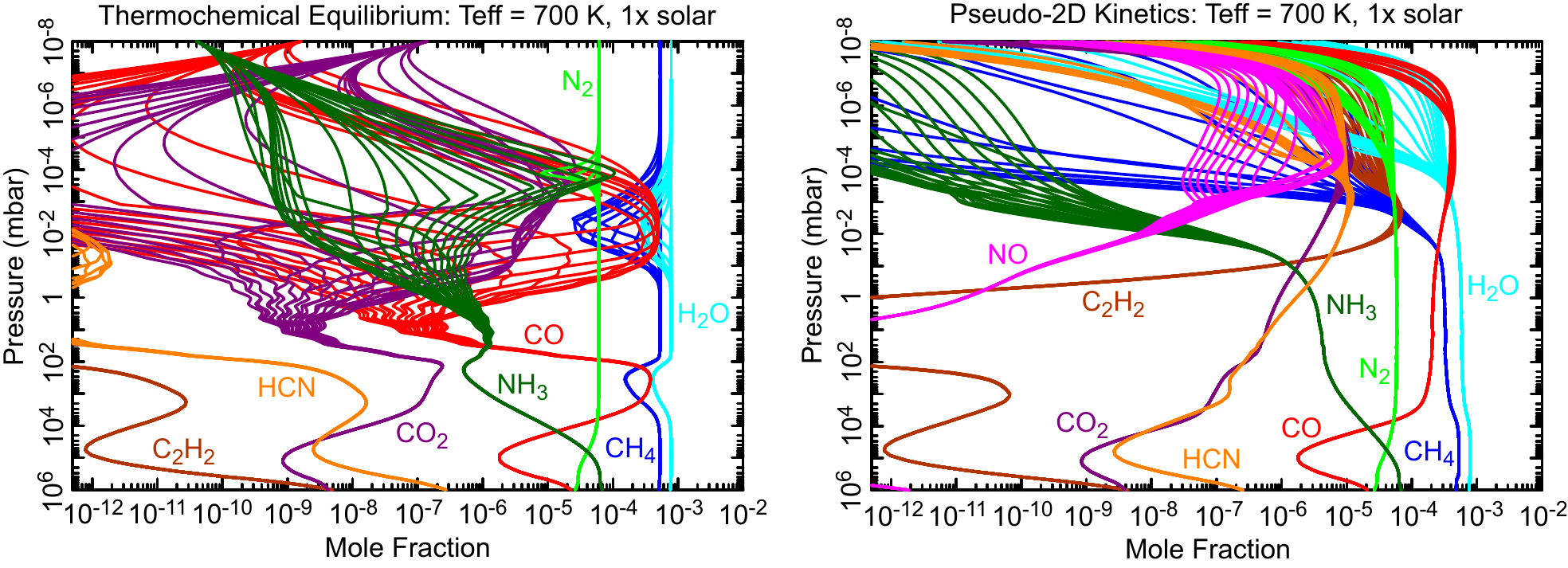} \\
\vspace{10pt}
\includegraphics[width=1.0\textwidth]{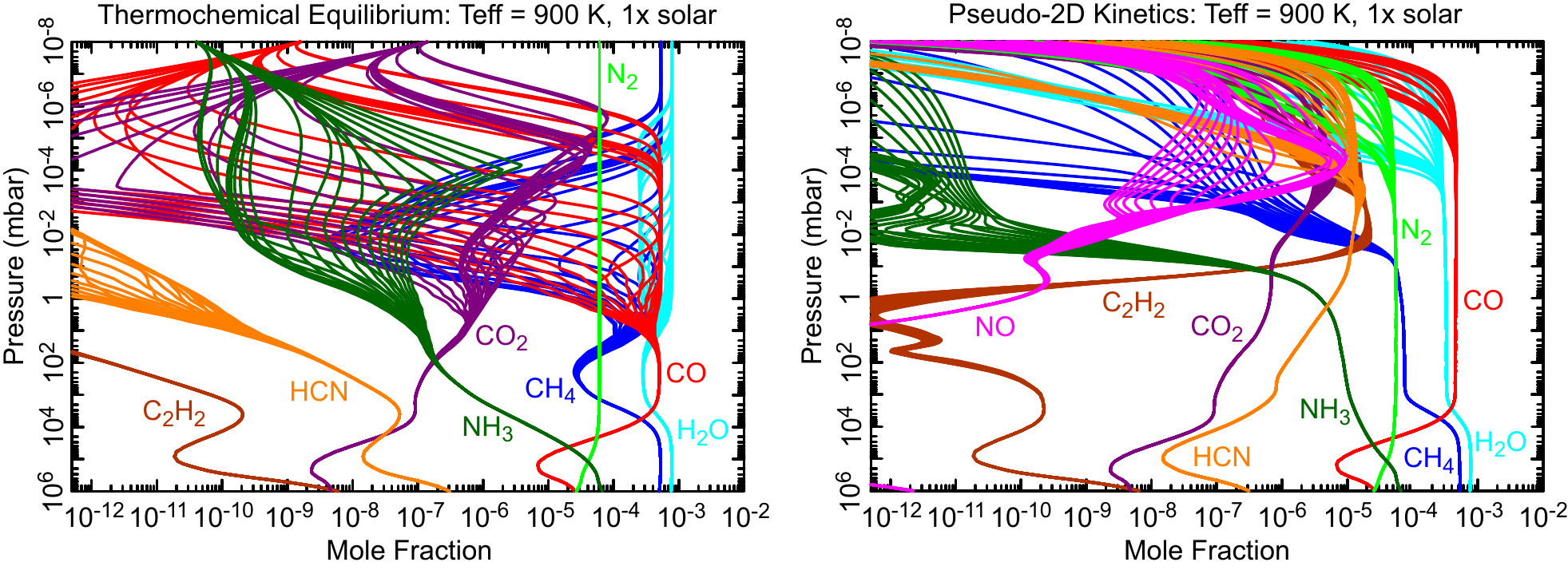} \\
\end{center}
\vspace{-0.5cm}
\noindent{Fig.~\ref{figmixall}. \textit{(continued)}.}
\end{figure*}

\clearpage

\begin{figure*}[!htb]
\begin{center}
\includegraphics[width=1.0\textwidth]{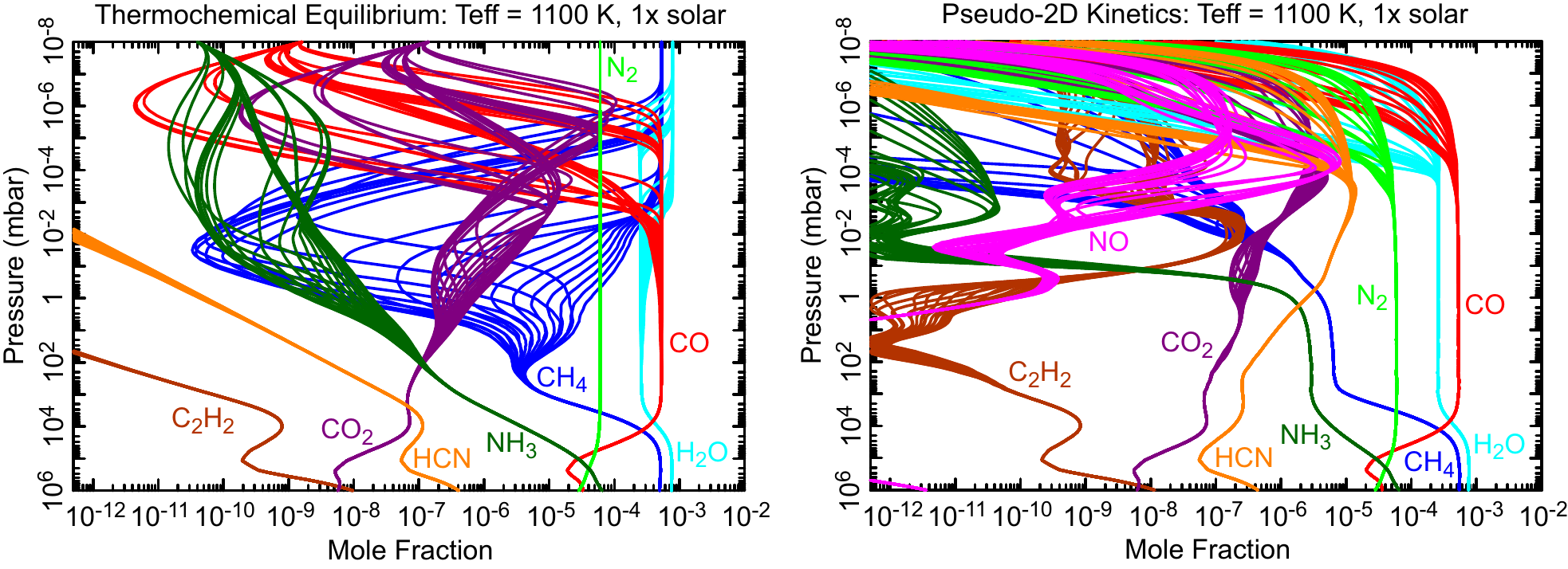} \\
\vspace{10pt}
\includegraphics[width=1.0\textwidth]{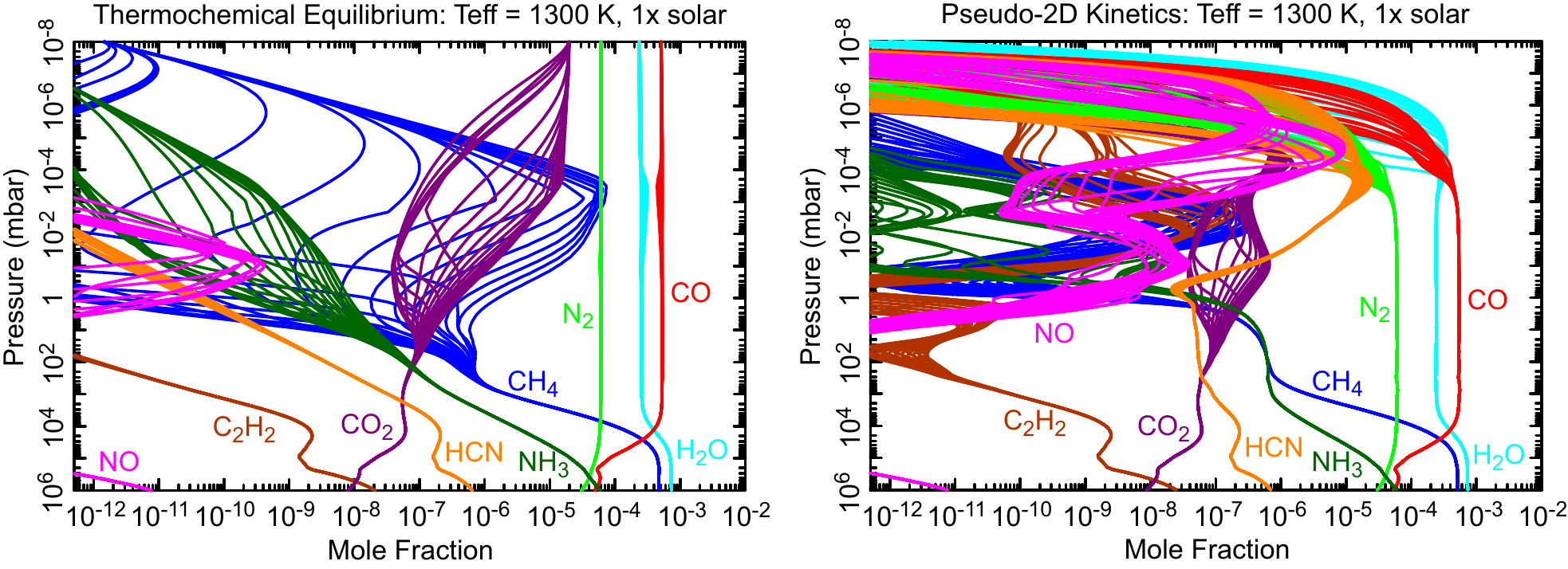} \\
\vspace{10pt}
\includegraphics[width=1.0\textwidth]{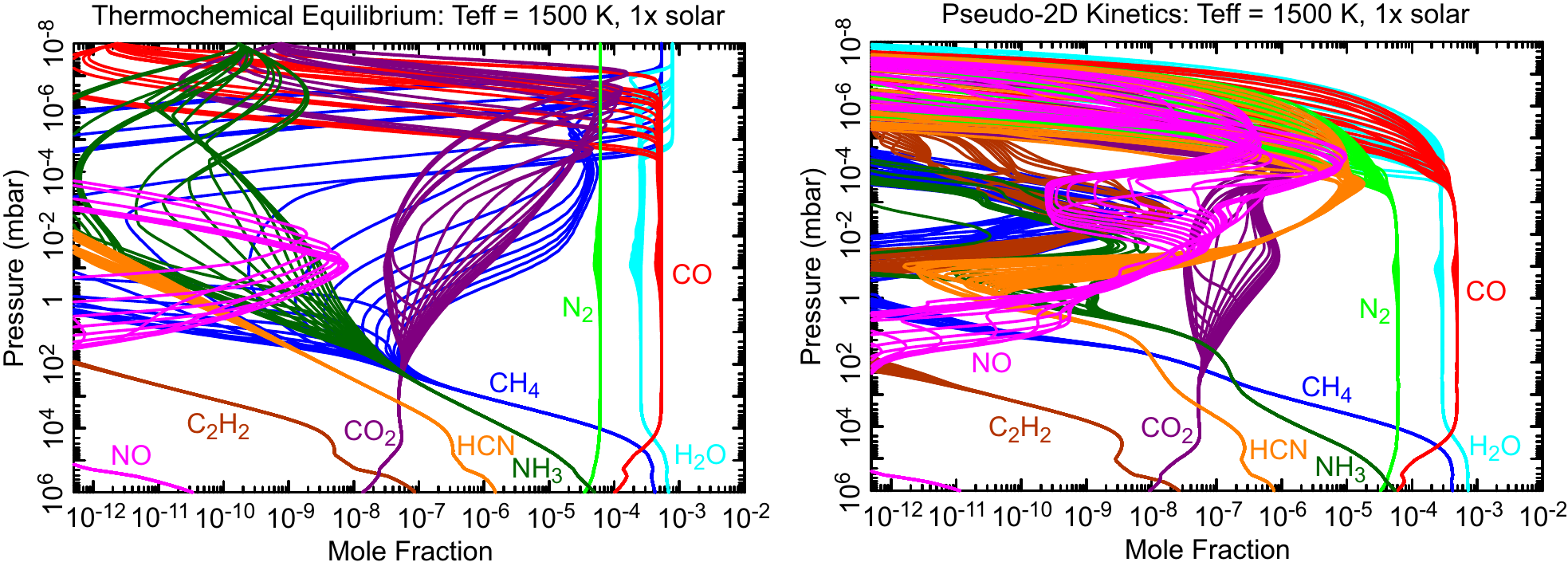} \\
\end{center}
\vspace{-0.5cm}
\noindent{Fig.~\ref{figmixall}. \textit{(continued)}.}
\end{figure*}

\clearpage

\begin{figure*}[!htb]
\begin{center}
\includegraphics[width=1.0\textwidth]{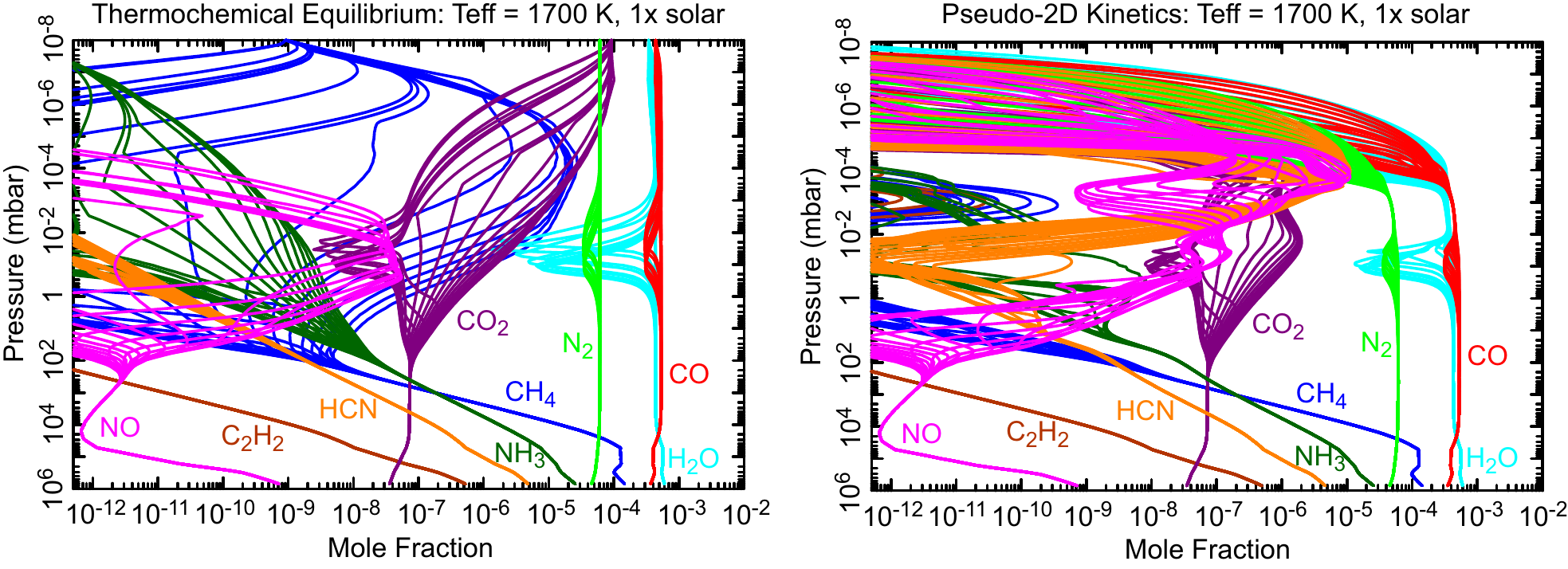} \\
\vspace{10pt}
\includegraphics[width=1.0\textwidth]{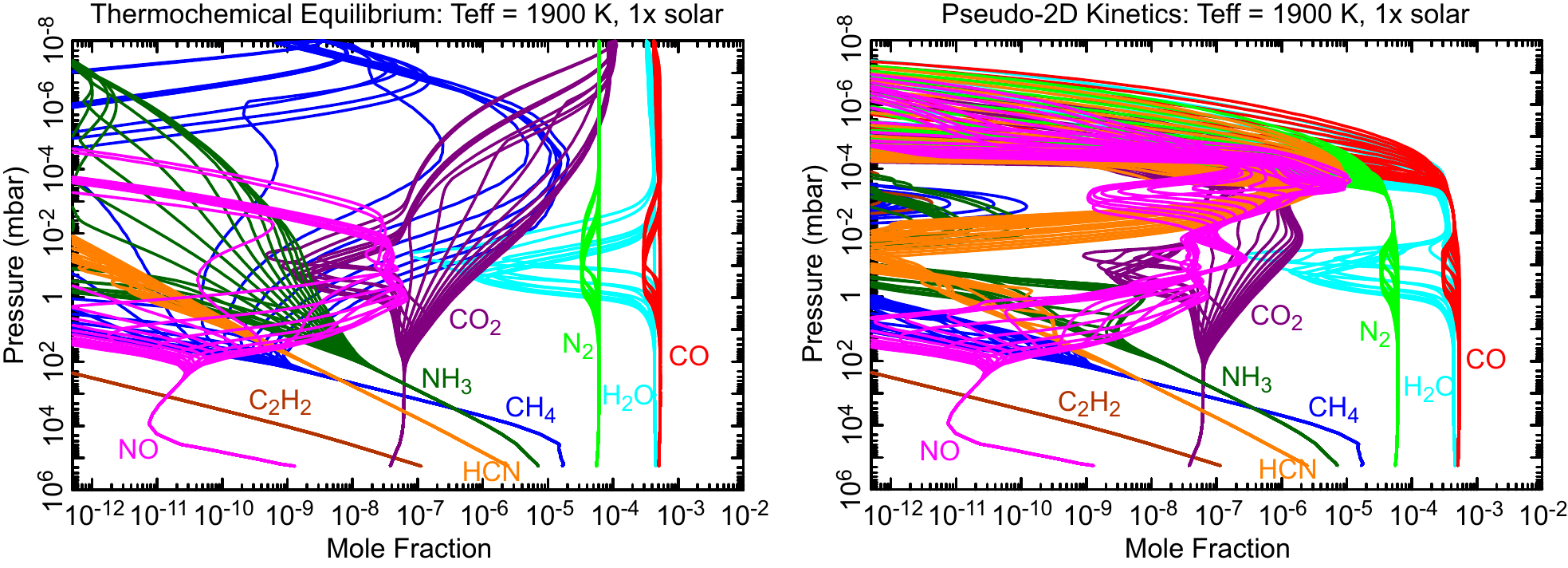} \\
\vspace{10pt}
\includegraphics[width=1.0\textwidth]{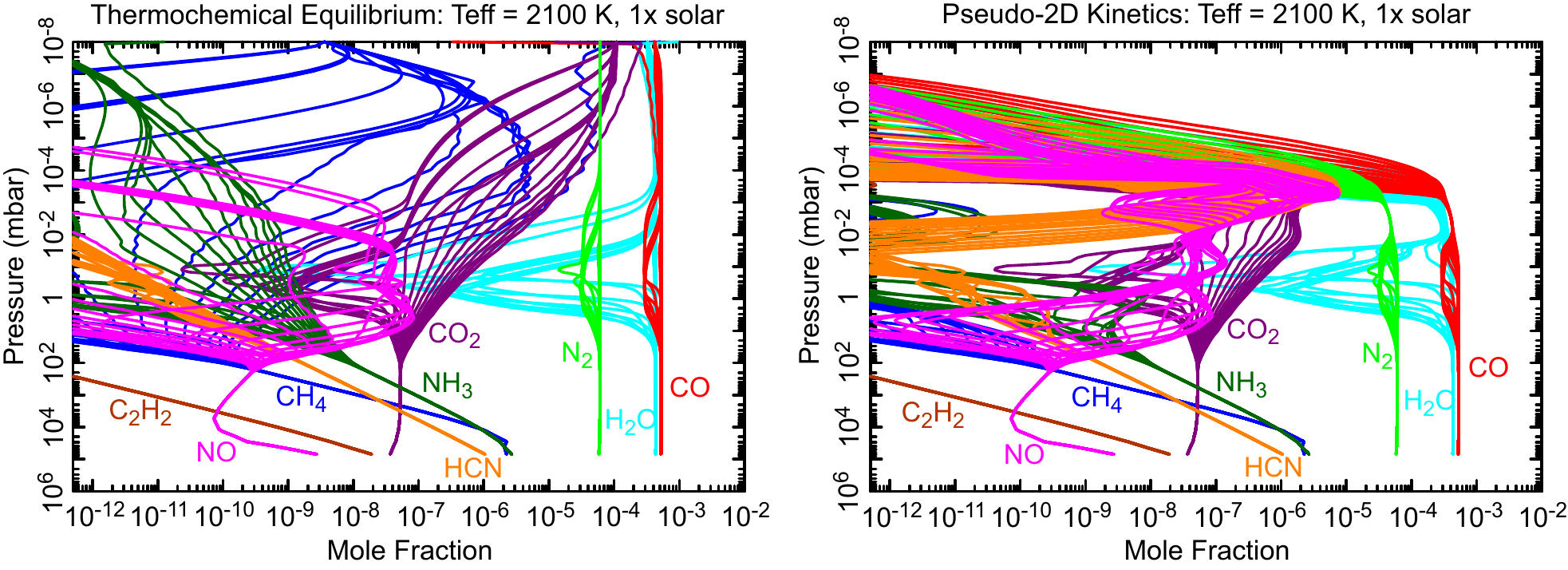} \\
\end{center}
\vspace{-0.5cm}
\caption{Mixing ratios for several key species in our grid of solar-composition exo-Neptune models, as a function of altitude and 
longitude: (Left) assuming thermochemical equilibrium, and (Right) predicted by our pseudo-2D chemical-kinetics models.  Each species is 
represented by a different color (as labeled), and vertical profiles for all 20 longitudes are shown for each species.  The planetary \teff\ is 
listed at the top of each image.} 
\label{figmixall}       
\end{figure*}

For the exo-Neptune with \teff\ = 500 K, neither the thermochemical-equilibrium model, nor the pseudo-2D kinetics model exhibit strong longitude 
variations in abundance of the major species in the pressure regions where the emission observations are most sensitive, from $\sim$0.1-1000 mbar.  For 
the equilibrium model, the lack of significant variation of major species with longitude is caused by the confinement of temperature variations to higher 
altitudes (see Fig.~\ref{figtemp}) and to the fact that the temperature profile across the globe is solidly in the CH$_4$-dominated region at all 
longitudes and altitudes, with CO then being a minor species.  In terms of the nitrogen species, the \teff\ = 500 K planet is mostly in the N$_2$-dominated 
region, except for a narrow altitude region near $\sim$3$\scinot-4.$ mbar at some night-side longitudes and in  the $\sim$2--100 mbar region, where 
NH$_3$ becomes dominant at all longitudes.  Variations in minor constituents such as NH$_3$ and CO in 
the thermochemical-equilibrium model are therefore relatively muted and are confined to higher altitudes.  In contrast, quenching of CH$_4$-CO-H$_2$O and 
NH$_3$-N$_2$ occurs in the pseudo-2D kinetics model for the \teff\ = 500 K planet, with the quench point occurring at pressures where CH$_4$ and N$_2$ 
are dominant.  Carbon monoxide and NH$_3$ are important quench species in that case, but rapid zonal winds keep their abundances constant with longitude.  CO is 
also a very important product from coupled CH$_4$ and H$_2$O photochemistry in this model and therefore becomes much more abundant in the stratosphere 
in the pseudo-2D kinetics model than the thermochemical-equilibrium model.  By the same token, chemical reactions allow rapid interconversion between 
H$_2$O, CO, and CO$_2$, causing CO$_2$ to maintain a pseudo-equilibrium with H$_2$O and the enhanced quenched abundance of CO, so that CO$_2$ is greatly 
enhanced in abundance in the pseudo-2D kinetics model compared with the equilibrium model.  Coupled photochemistry of NH$_3$ and CH$_4$ produces HCN 
\cite{moses11}, 
which is also enhanced many orders of magnitude in the pseudo-2D model compared with the equilibrium model.  Some variations in species abundance 
with longitude do occur in the pseudo-2D model from variations in stellar flux with longitude, but such variations are confined to high altitudes.

The results for the \teff\ = 700 K planet are similar to those at \teff\ = 500 K.  Variations with longitude become more evident in the thermochemical 
equilibrium model because diurnal temperature variations are larger, and the thermal structure crosses back and forth between the 
regions where CH$_4$ and CO are the dominant carbon constituents, causing the abundance of these species to change more significantly with longitude.  
Carbon dioxide also becomes a more important minor constituent under these conditions.  Some diurnal variations in H$_2$O show up near 10$^{-2}$--10$^{-3}$ 
mbar as CO takes up more of the oxygen on the day side of the planet.  Because horizontal quenching by the rapid zonal winds is effective, the pseudo-2D model 
does not exhibit these diurnal variations in the observable regions of the stratosphere, although variations still exist at higher altitudes.  
Quenching and rapid kinetics between some quenched species allow CO, CO$_2$, and HCN to be important constituents of the pseudo-2D atmosphere.

The temperature structure of the \teff\ = 900 K planet straddles the CO = CH$_4$ equal abundance line (see Fig.~\ref{figtemp}), causing large 
swings in the diurnal variation of the atmospheric composition in the thermochemical-equilibrium model.  At this \teff , CO and CO$_2$ become much more prominent 
during the day; CH$_4$, NH$_3$, and H$_2$O are more abundant on the cooler night side.  Thanks to effective horizontal quenching, the pseudo-2D 
kinetics model has fewer diurnal variations in composition, but the atmospheric temperatures are now warm enough that the CH$_4$-CO-H$_2$O quench 
point is in the CO-dominated regime.  Carbon monoxide is therefore more abundant than CH$_4$ in the observable region of the atmosphere, and CO 
even becomes more abundant than H$_2$O.  The quenched abundance of NH$_3$ is now greater than that in the lower stratosphere of the thermochemical 
equilibrium model, which was not the case with the cooler planets.  The higher incident stellar flux on this \teff\ = 900 K planet causes more 
photochemical destruction of CO and N$_2$ at very high altitudes, and some of the carbon and nitrogen released from this photochemistry ends up 
producing HCN, which is still a key photochemical product, along with NO and C$_2$H$_2$.  

At \teff\ = 1100 K, CO becomes the dominant carbon constituent throughout the observable regions of the stratosphere in the thermochemical 
equilibrium model, but CH$_4$ still exhibits significant diurnal variation, being more abundant in the nighttime hemisphere.  In fact, CH$_4$ 
is more abundant on the night side in the equilibrium model than it is globally in the pseudo-2D model.  Diurnal variations in CO$_2$ and NH$_3$ 
are also present down to fairly deep stratospheric pressures, whereas diurnal variations in CO and H$_2$O are more prominent at high altitudes in 
the equilibrium model.  Again, there is less diurnal variation in general in the pseudo-2D \teff\ = 1100 K model than in thermochemical 
equilibrium, except at high altitudes.  The quench points for this warmer planet are further into the CO and N$_2$ stability fields, leading to 
smaller quenched abundances of CH$_4$ and NH$_3$ than the cooler planets.  The photochemical production of hydrocarbons is correspondingly reduced
in the middle stratosphere.  

Moving inward in orbital distance to the \teff\ = 1300 and 1500 K planets, CO, H$_2$O, and N$_2$ become firmly entrenched as the most abundant 
stratospheric constituents after H$_2$ and He in both the thermochemical-equilibrium and pseudo-2D kinetics models.  These three species exhibit 
little variation with longitude in the stratosphere in either model.  CO$_2$ and CH$_4$ are still important stratospheric constituents in both 
models, with some diurnal variation seen in both (with CH$_4$ being more abundant on the cooler night side, CO$_2$ more abundant on the warmer day side).  
Quenched CH$_4$ becomes much less of a factor in the pseudo-2D model, especially at 1500 K, and the overall composition more closely tracks thermochemical 
equilibrium.  Photochemical products are still present, especially at high altitudes, but are significantly reduced in the middle and lower 
stratosphere due to higher temperatures and more efficient conversion to a thermochemical-equilibrium state.

At \teff\ $\ge$ 1700 K, stratospheric temperature variations across the planet are very large (see Fig.~\ref{figtemp}).  Day-side temperatures become 
large enough to affect H$_2$O, which becomes depleted in the middle atmosphere in the daytime but recovers at night.  The pseudo-2D kinetics model 
results more closely follow the thermochemical equilibrium results at these higher temperatures, except that horizontal quenching causes somewhat 
reduced diurnal variations in H$_2$O and CO.  Longitudinal variations in stratospheric CO$_2$ are important in both models.  Methane is no longer 
an important constituent in the higher-temperature planets in the pseudo-2D models, but CH$_4$ is still present at nighttime in the 
thermochemical-equilibrium model.  Atomic species and small radicals are the dominant photochemical products, especially at high altitudes. 

\subsection{Phase-curve emission results} \label{sec:phasecurves}

Figure ~\ref{figphase} illustrates the effects of the predicted altitude- and longitude-variable temperatures and composition on the emission 
spectra and phase curves of our grid of exo-Neptunes at visible and infrared wavelengths.  The emission spectra are shown in the left 
column and the phase curves at specific wavelengths are shown in the right column.  A phase angle of zero in these figures corresponds to eclipse 
geometry with the fully illuminated day-side hemisphere facing the observer.

The emission predictions in Fig.~\ref{figphase} do not include various sources of noise that are expected to affect \textit{Ariel} phase-curve observations 
(see Mugnai et al. \cite{mugnai20} for a full discussion of noise sources and a description of their physically motivated noise model for the mission, 
\texttt{ArielRad}).  As is described in Mugnai et al. \cite{mugnai20}, photon noise is expected to dominate measurements with the \textit{Ariel} Infrared Spectrometer 
(AIRS) instrument, both in channel 0 (1.9--3.9 $\mu$m) and in channel 1 (3.9--7.8 $\mu$m) that will be critical for extracting the most science possible 
out of \textit{Ariel} phase-curve observations.  Realistic noise estimates depend significantly on target properties such as distance from the Earth and position 
on the sky, as well as specific observing plans that are not defined for these generic planets; therefore, we have not utilized \texttt{ArielRad} or other realistic 
noise simulators.  Instead, we use the more realistic \textit{Ariel} phase-curve simulations for specific planetary targets provided by Charnay et al. \cite{charnay20} 
as a guide for qualitatively evaluating the detectability of our predicted phase curves with \textit{Ariel}.  Charnay et al. \cite{charnay20} describe that the 
general requirement for spectroscopic phase curves is to reach a signal-to-noise ratio of greater than 10 at a resolving power $R$ $\approx$ 50 (e.g., 
relevant to AIRS channel 1).  They present example simulations that demonstrate that peak-to-trough phase-curve amplitude variations of $\gta$10$^{-4}$ (in 
the flux of the planet divided by the flux of the star) are detectable for the sub-Neptune GJ 1214 b from one orbit observed with AIRS channel 1 
(see Fig.~1 of \cite{charnay20}).  We therefore consider amplitudes of $\sim$10$^{-4}$ or greater as the detection threshold for our phase-curve predictions.

\begin{figure*}[!htb]
\begin{center}
\includegraphics[width=1.0\textwidth]{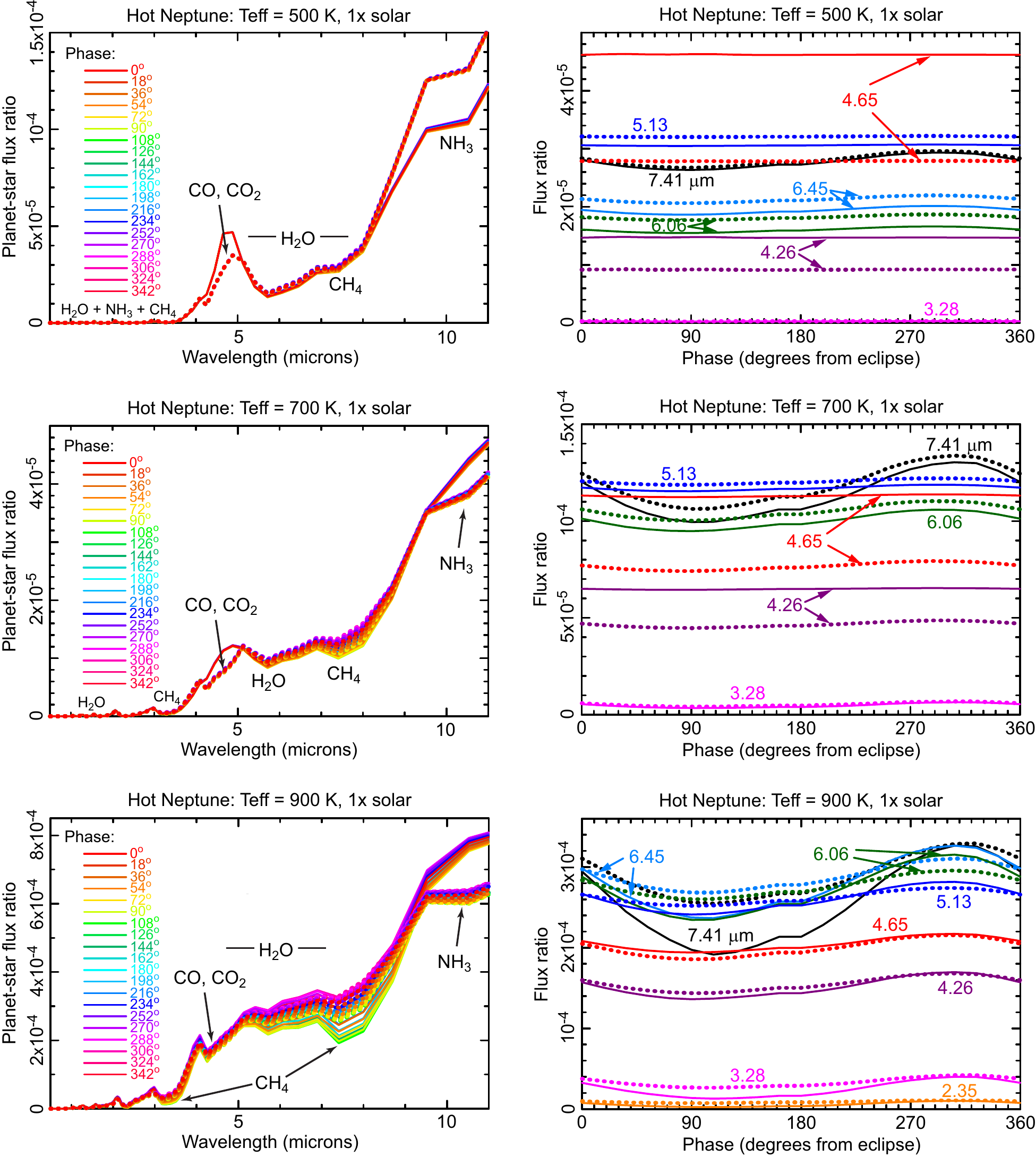} 
\end{center}
\noindent{Fig.~\ref{figphase}. \textit{(continued)}.}
\end{figure*}

\clearpage

\begin{figure*}[!htb]
\begin{center}
\includegraphics[width=1.0\textwidth]{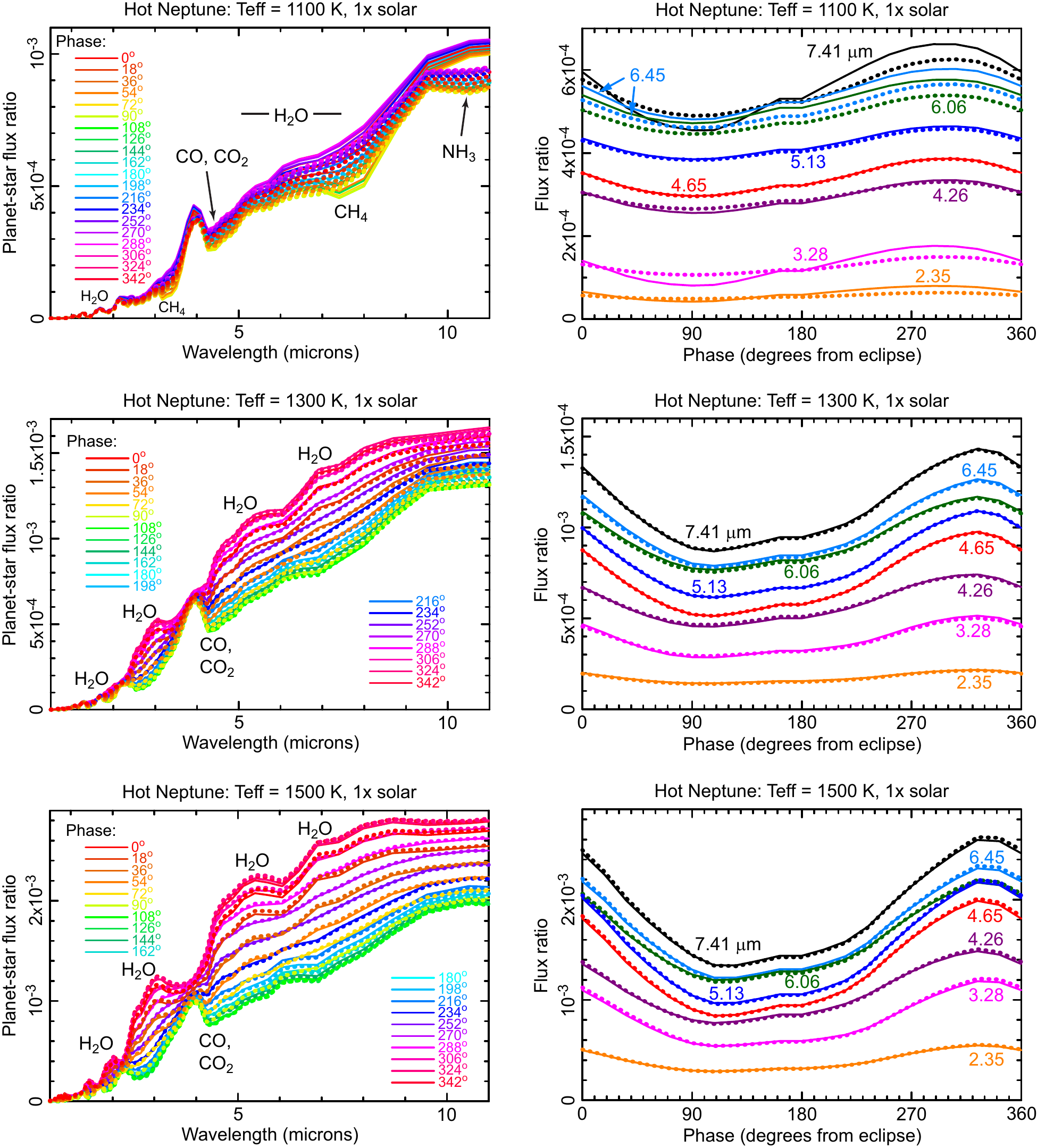} 
\end{center}
\noindent{Fig.~\ref{figphase}. \textit{(continued)}.}
\end{figure*}

\clearpage

\begin{figure*}[!htb]
\begin{center}
\includegraphics[width=1.0\textwidth]{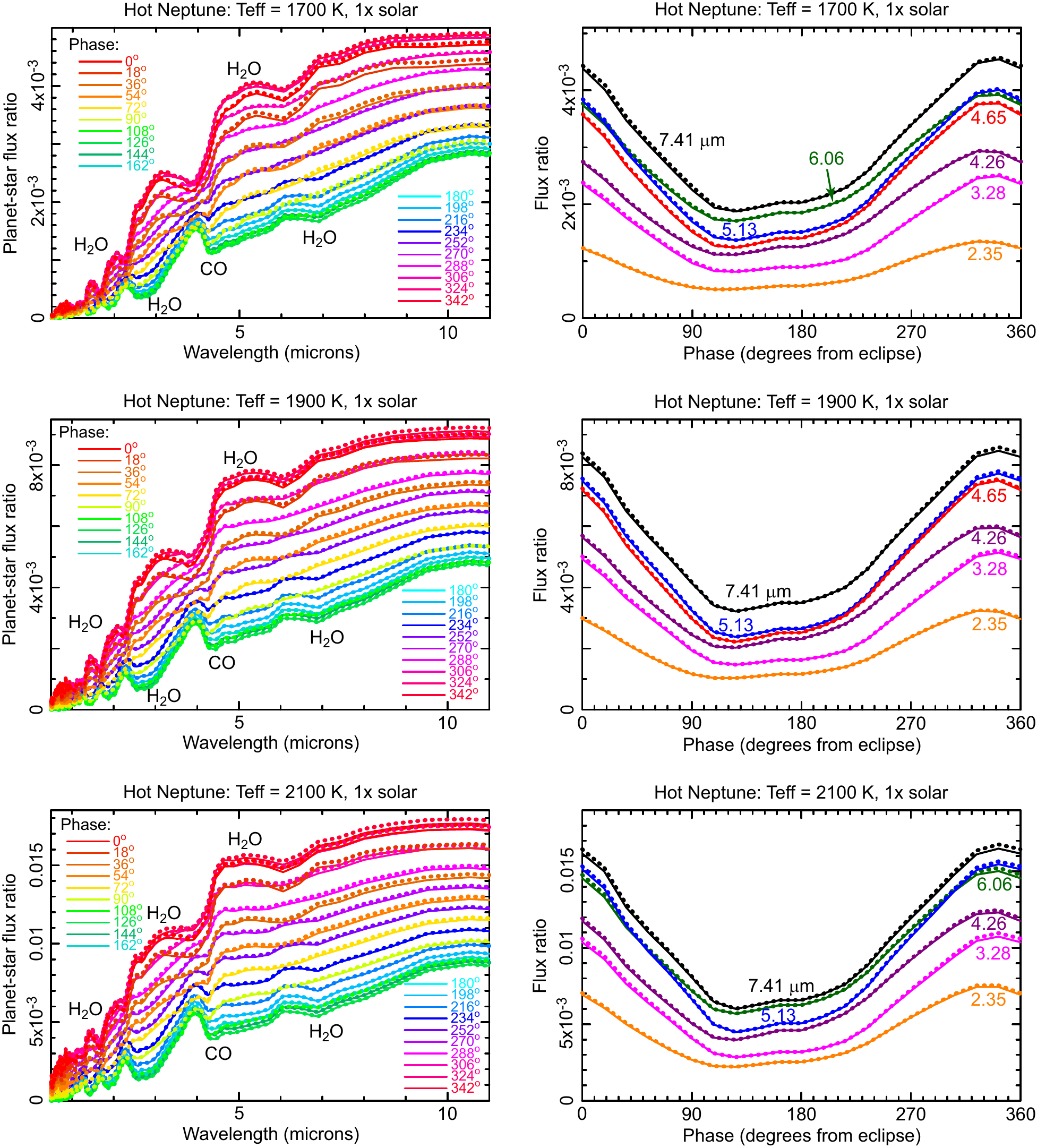}
\end{center}
\vspace{-0.5cm}
\caption{Emission flux from the planet divided by the emission flux from the star: (Left) as a function of wavelength over the planet's 
orbit, with different orbital phase angles (starting from eclipse geometry defined as 0$^{\circ}$ phase) in different colors, and 
the predictions from the thermochemical-equilibrium model depicted as solid curves, with the pseudo-2D kinetics model as dotted curves;    
(Right) as a function of orbital phase at specific wavelengths (color coded, as labeled), starting from eclipse at 0$^{\circ}$ phase.  Different 
rows correspond to the different planetary \teff\ models, as labeled at the top of the figures.  Note 
that the actual planetary transits and eclipses are omitted from the phase curves.}
\label{figphase}       
\end{figure*}

The emission phase curves for the \teff\ = 500 K planet are relatively flat with planetary phase and unlikely to be detectable with \textit{Ariel}.  Diurnal 
temperature variations at the 
pressures where the emission originates ($\sim$0.1-1000 mbar) are small, and the atmospheric composition does not change much with longitude 
at these pressures for either the thermochemical-equilibrium or pseudo-2D kinetics model.  The very small amplitude phase-curve differences 
that can be seen at 6.06--7.41 $\mu$m in the \teff\ = 500 K phase-curve figure are due to temperature variations with longitude in the $\sim$0.1-1 mbar 
region, and not to variations in composition.  However, the differences in composition between the thermochemical-equilibrium and kinetics models 
do show up as notable offsets in the phase curves at certain wavelengths, such as 4.26 and 4.65 $\mu$m, where CO$_2$ and CO, respectively, have 
absorption bands.  The pseudo-2D kinetics model exhibits greater absorption at these wavelengths due to the much greater abundance of 
quenched CO and the associated kinetically produced CO$_2$.  The greater abundance of CO and CO$_2$ in the pseudo-2D kinetics model leads to emission 
from higher altitudes, where temperatures are lower, leading to less flux in the CO and CO$_2$ bands; this difference in emission shows up clearly 
in the actual emission spectrum at $\sim$4.2-5 $\mu$m.  
Conversely, at longer wavelengths, the greater abundance of NH$_3$ in the 0.1-10 mbar region of the thermochemical-equilibrium model leads to greater 
apparent absorption in NH$_3$ bands near 6.1 $\mu$m in the equilibrium model, and again more prominently in a broad region centered near 10.5 $\mu$m; i.e., 
the greater abundance of NH$_3$ in the 0.1-10 mbar region for the thermochemical equilibrium model causes the emission at the wavelengths within the NH$_3$ 
vibrational bands to originate from higher altitudes, where the temperatures are lower, resulting in less emission at these wavelengths.
Methane and water bands are obvious in 
the spectra from both models, but the similar CH$_4$ and H$_2$O abundances in the two models leads to similar spectral signatures in molecular 
bands from these species.

At \teff\ = 700 K, the spectral behavior is qualitatively similar to the 500 K case.  Differences in the 4-5 $\mu$m emission between the equilibrium 
and kinetics models still appear due to the greater abundance of CO and CO$_2$ in the pseudo-2D kinetics model, but these differences are proportionally 
smaller due to the increase in the equilibrium CO abundance, in particular.  More significant differences in the phase-curve amplitudes start to 
appear at 6-9 $\mu$m near the peak of the blackbody function due to stronger temperature variations with longitude.  The CH$_4$ absorption band 
centered near 7.8 $\mu$m becomes more obvious, as does its variation with atmospheric temperature/longitude.  At $\sim$9.5-11 $\mu$m, the equilibrium and 
kinetics models have notably different absorption, but this time it is the pseudo-2D model that has greater absorption in this region due to the 
greater quenched abundance of NH$_3$.  Overall, variations in emission with planetary phase in the \teff\ = 700 K model are small and unlikely to be 
detectable at wavelengths relevant to \textit{Ariel}.  

Phase-curve amplitudes in the 6-9 $\mu$m wavelength region increase for the \teff\ = 900 K model, again aided by larger temperature changes with 
longitude, although the different diurnal behavior of CH$_4$ between the equilibrium and kinetics models becomes obvious at this \teff .  The very 
large swings in the CH$_4$ abundance with longitude in the thermochemical-equilibrium model lead to large amplitude changes in the emission as 
a function of planetary phase in the CH$_4$ band centered near 7.8 $\mu$m.  In the pseudo-2D kinetics model, the quenched CH$_4$ abundance stays mostly 
constant with longitude in the pressure region responsible for the emission, resulting in a flatter phase curve at these wavelengths.  Similar 
differences between the equilibrium and kinetics model also appear in the 3.3 $\mu$m band of CH$_4$, although there is overall less emission and 
less variation with planetary phase at these shorter wavelengths.  In fact, the phase curves are flatter with the pseudo-2D kinetics model at many 
wavelengths, due to the more uniform abundances as a function of longitude.  The kinetics model contains more NH$_3$ as a result of quenching, so it 
exhibits more absorption in the 6.1 and 10.5 $\mu$m wavelength regions.  On the other hand, now that CO is the dominant carbon constituent at 
$P$ $>$ 10 mbar in both models --- which also leads to similar CO$_2$ abundances, as CO$_2$ remains in a pseudo-equilibrium with CO and H$_2$O in 
the kinetics model --- the differences in emission in the 4.2-5 $\mu$m regions between the two models largely disappears at planetary \teff\ = 900 K.  
Our modeling suggests that if the atmosphere remains in thermochemical equilibrium throughout the 1$\times$ solar metallicity, \teff\ = 900 K exo-Neptune, 
then the phase-curve amplitudes in the methane band centered near $\sim$7.8 $\mu$m would be detectable with \textit{Ariel}, but would not be detectable at other 
wavelengths or if disequilibrium chemistry were to dominate the atmospheric composition.

Increasing the planetary \teff\ further, to 1100 K, results in larger temperature swings with longitude, resulting in larger phase curve amplitudes 
that extend to shorter  wavelengths.  Again, the larger day-night swings in the CH$_4$ abundance in the thermochemical equilibrium model lead to greater 
amplitudes in the phase curves at wavelengths where CH$_4$ has bands, including $\sim$2.3, 3.3, and 7.8 $\mu$m.  The greater NH$_3$ abundance due to 
quenching in the pseudo-2D kinetics model leads to greater absorption in the 6.1 and 10.5 $\mu$m bands.  Water, CO, and CO$_2$ have similar abundances 
in the two models, resulting in similar emission at wavelengths where those species dominate and nearly identical phase curves at those wavelengths.  
Phase-curve amplitudes are more likely to be detectable with \textit{Ariel}, particularly if the atmosphere remains in thermochemical equilibrium.

At \teff\ $\ge$ 1300 K, the composition in the pseudo-2D kinetics model more closely tracks thermochemical equilibrium, due to rapid kinetics and the 
strong dominance of CO, H$_2$O, and N$_2$ as the main carriers of O, C, and N.  The emission spectrum and phase curves of the two models are then 
quite similar.  The phase-curve amplitudes for both the disequilibrium and equilibrium cases in the \teff\ = 1300 K planet would be detectable within the 
AIRS channel 1.  The phase curve amplitudes continue to increase with increasing \teff , becoming detectable within both AIRS channels at \teff\ = 1500 K 
and hotter.  Molecular bands clearly switch from being 
seen in absorption on the cooler night side (where temperatures decrease with height) to emission on the hotter day side (where temperature inversions are 
present).  Note that the maximum in emission for all planetary \teff 's is not located exactly at 0$^{\circ}$ phase, as a result of the eastward offset of 
the hot-spot location.  The combination of the eastward zonal jet and finite radiative time constants causes the hottest temperatures on the planet to be 
shifted a few tens of degrees in eastward in longitude from the substellar point \cite{knut07,showman09}.  

\section{Sensitivity to metallicity} \label{sec:sensitivity}

The results from section \ref{sec:results} show how the atmospheric temperature structure, composition, and emission behavior change systematically 
as planetary \teff\ is increased for solar-composition atmospheres.  Although Neptune-class exoplanets are expected to have H$_2$-rich atmospheres, 
their complement of heavy elements is likely to be greater than solar, both because they most likely formed originally 
with higher-metallicity atmospheres \cite{fort13frame,helled14uran} and because smaller exoplanets that reside close to their host stars will be 
subject to strong extreme-ultraviolet irradiation that will substantially heat their upper atmospheres, leading to the loss of H and He over time 
\cite{garcia07,murrayclay09,owen13,lopez13,lopez14}.  We therefore examine the sensitivity of the model results to atmospheric metallicity. Specifically,  
we examine the results for an enhanced metallicity of 100$\times$ solar, leaving all other planetary properties (1-bar radius, mass, orbital distance, 
zonal wind speed, $K_{zz}$) the same as with our 1$\times$ solar models, although we should note that both the zonal wind velocity and $K_{zz}$ are expected 
to have some sensitivity to the atmospheric mean molecular mass \cite{kataria15,lewis10,showman10,zhang17}.  Uranus and Neptune have a carbon metallicity of 
$\sim$60-130$\times$ solar \cite{sromovsky14,irwin19nepch4}, so the assumption of 100$\times$ solar metallicity for our exo-Neptunes planets is not 
unreasonable.  We restrict our grid here to ``warm'' planets.  Cold planets will have low-amplitude phase curves (see section \ref{sec:phasecurves}) 
and are therefore risky targets for \textit{Ariel} phase-curve observations, whereas hot atmospheres are dominated by thermochemical 
equilibrium chemistry (see section \ref{sec:chemresults}), leading to few differences between kinetics and equilibrium models.  Very hot Neptunes 
may have lost their entire atmospheric envelope through atmospheric escape (e.g., the so-called ``Neptune desert'' in exoplanet population 
statistics, see \cite{kurokawa14,sanchisojeda14,mazeh16}).
 
\begin{figure*}[!htb]
\begin{center}
\includegraphics[width=0.49\textwidth]{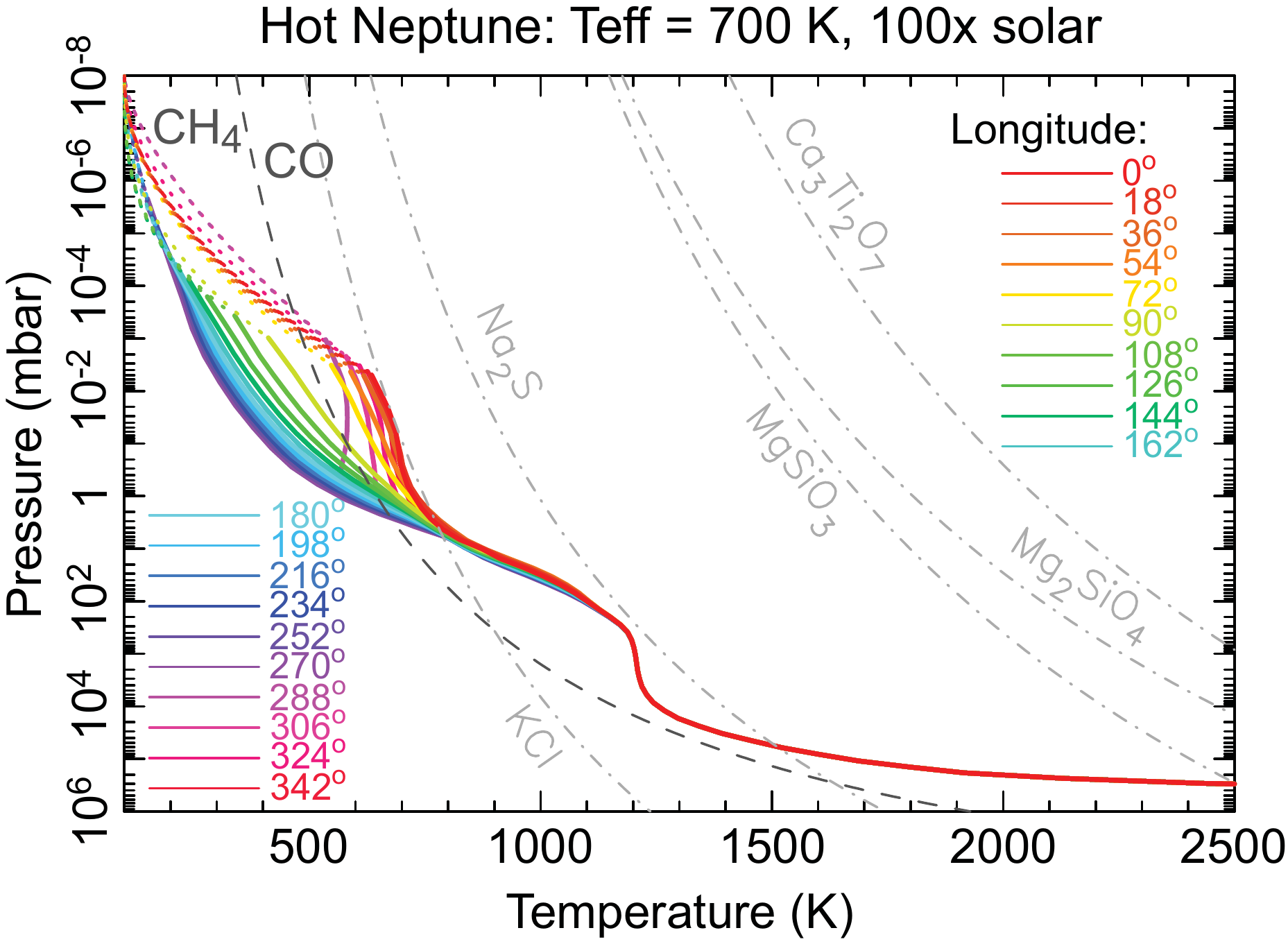} 
\includegraphics[width=0.49\textwidth]{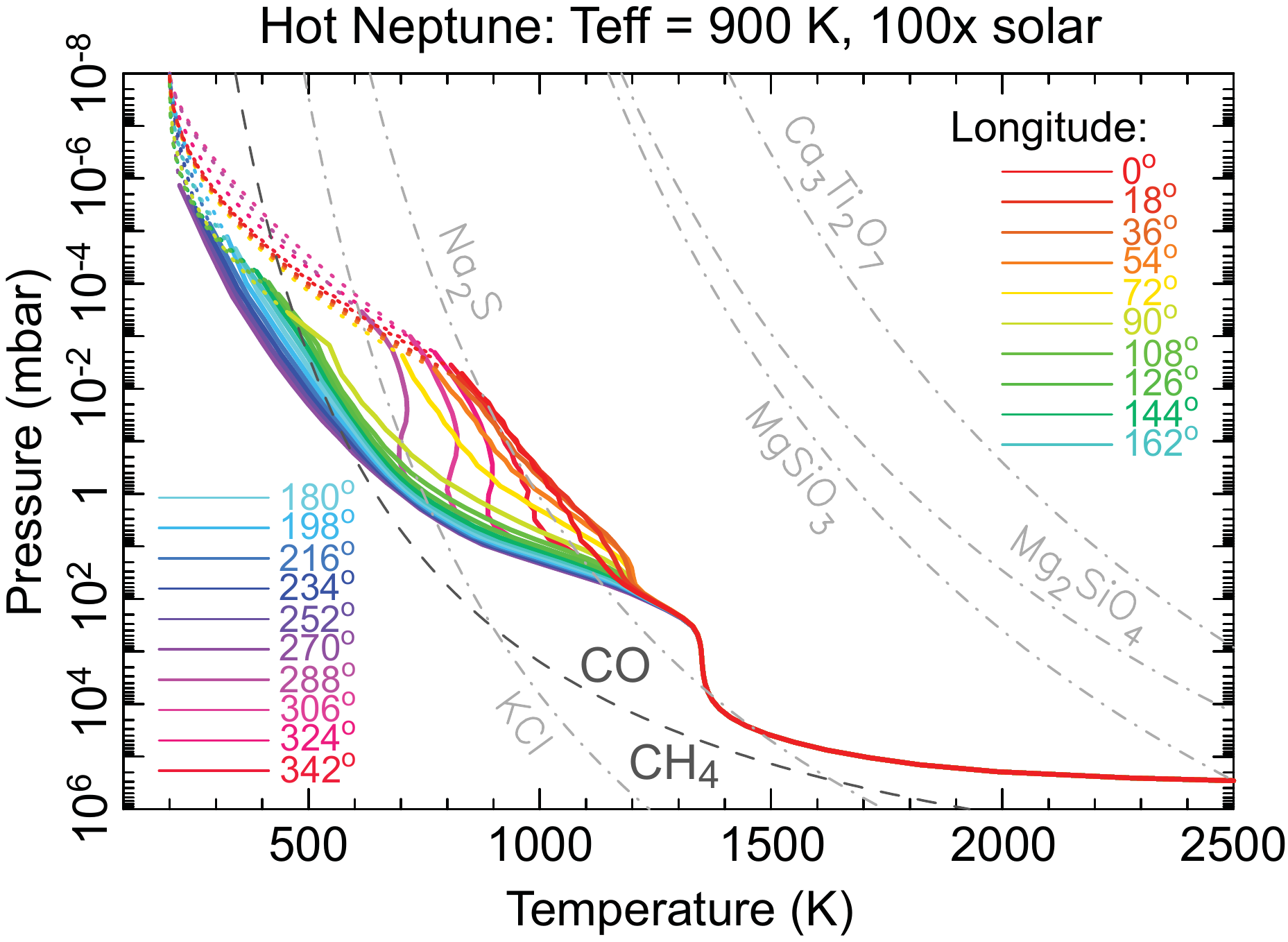} \\
\vspace{10pt}
\includegraphics[width=0.49\textwidth]{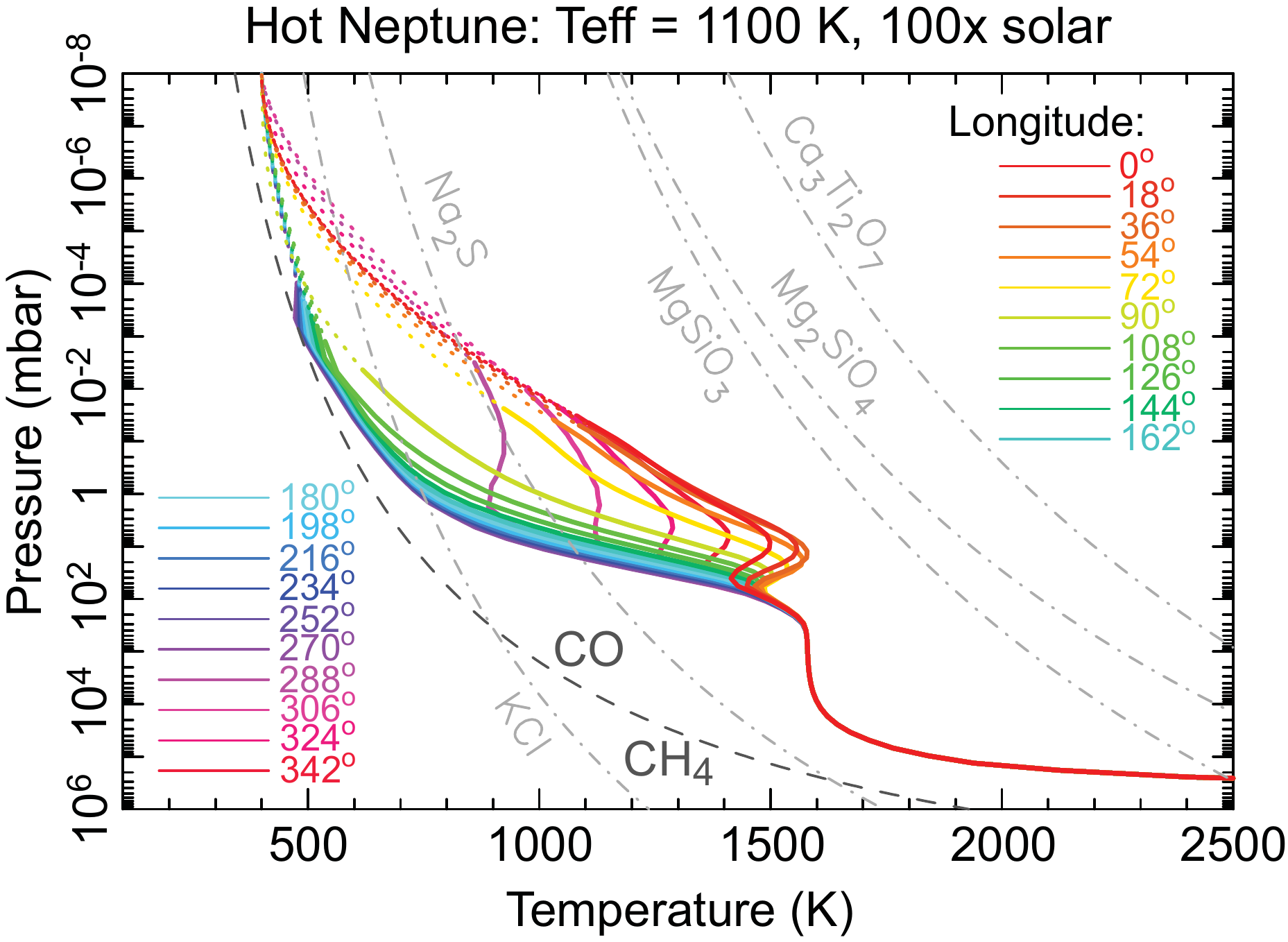}
\end{center}
\vspace{-0.5cm}
\caption{Equatorial temperature-pressure profiles as a function of longitude (colored solid lines, as labeled) for \teff\ = 700, 900, and 1100 K 
exo-Neptunes for an assumed 100$\times$ solar atmospheric metallicity.  See Fig.~\ref{figtemp} for further details.}
\label{figtemp100x}       
\end{figure*}

\begin{figure*}[!htb]
\begin{center}
\includegraphics[width=1.0\textwidth]{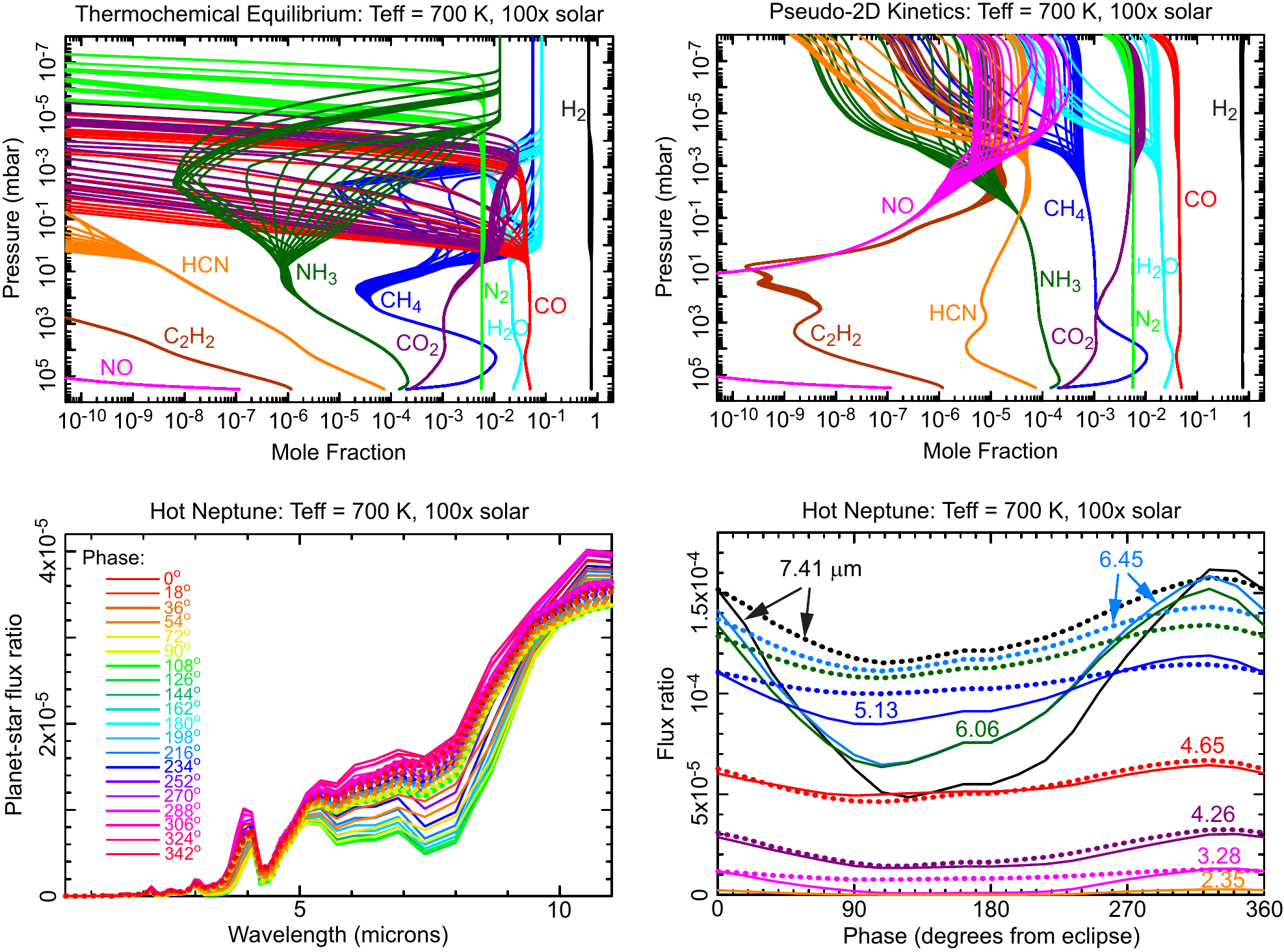} 
\end{center}
\vspace{-0.5cm}
\caption{Results from our modeling of the \teff\ = 700 K, 100$\times$ solar metallicity exo-Neptune: (Top Left) Vertical mixing ratio 
profiles for several key species at all longitudes in the thermochemical-equilibrium model; (Top Right) same as the Top Left, but for the 
pseudo-2D kinetics model; (Bottom Left) emission spectrum as a function of orbital phase (starting from eclipse geometry at 0$^{\circ}$) for 
the thermochemical-equilibrium model (solid curves) and pseudo-2D kinetics model (dotted curves); (Bottom Right) emission as a function of 
orbital phase at specific infrared wavelengths.}
\label{fig700p2}       
\end{figure*}

\begin{figure*}[!htb]
\begin{center}
\includegraphics[width=1.0\textwidth]{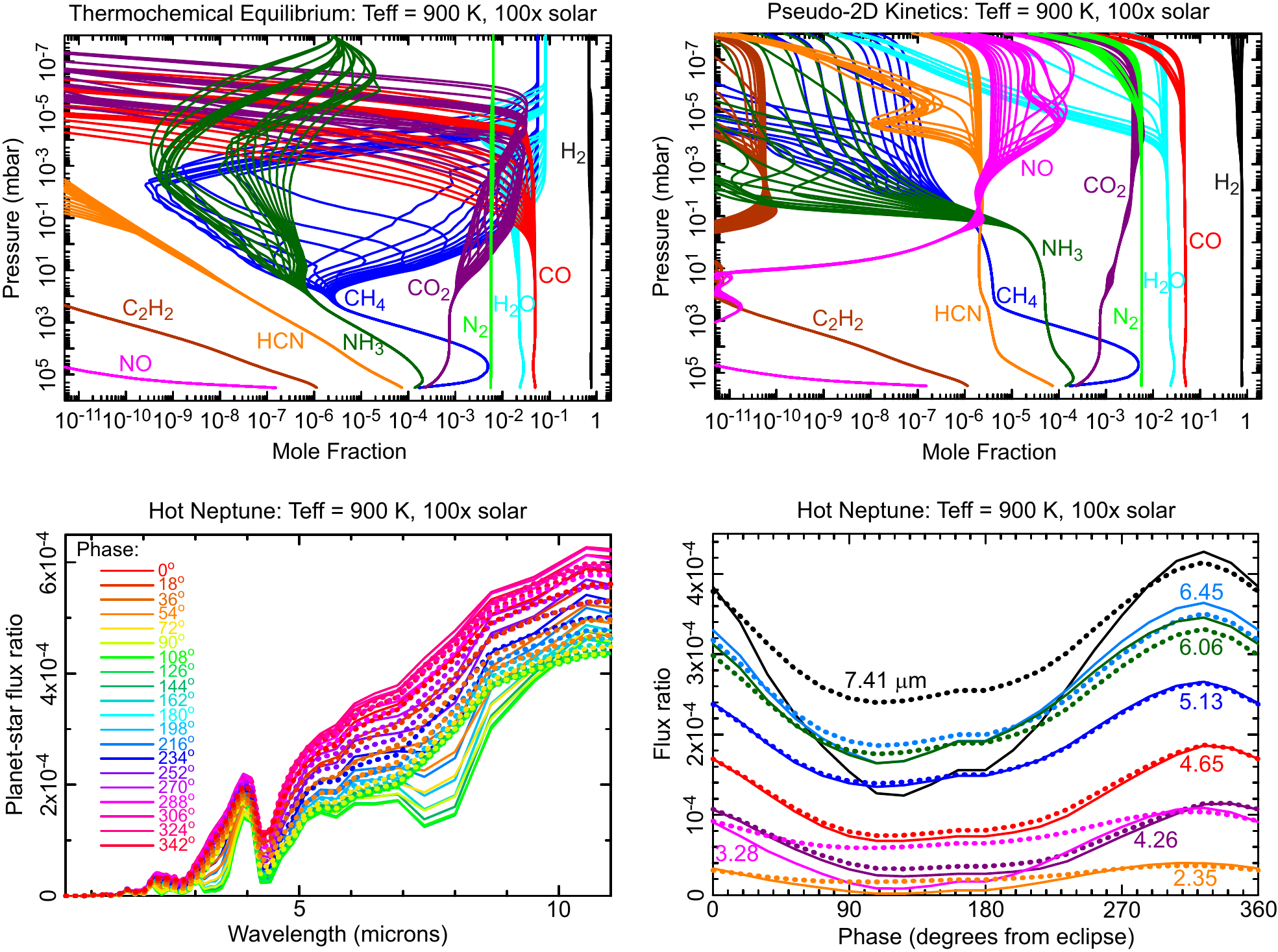} 
\end{center}
\vspace{-0.5cm}
\caption{Same as Fig.~\ref{fig700p2}, except for the \teff\ = 900 K, 100$\times$ solar metallicity exo-Neptune.}
\label{fig900p2}       
\end{figure*}

\begin{figure*}[!htb]
\begin{center}
\includegraphics[width=1.0\textwidth]{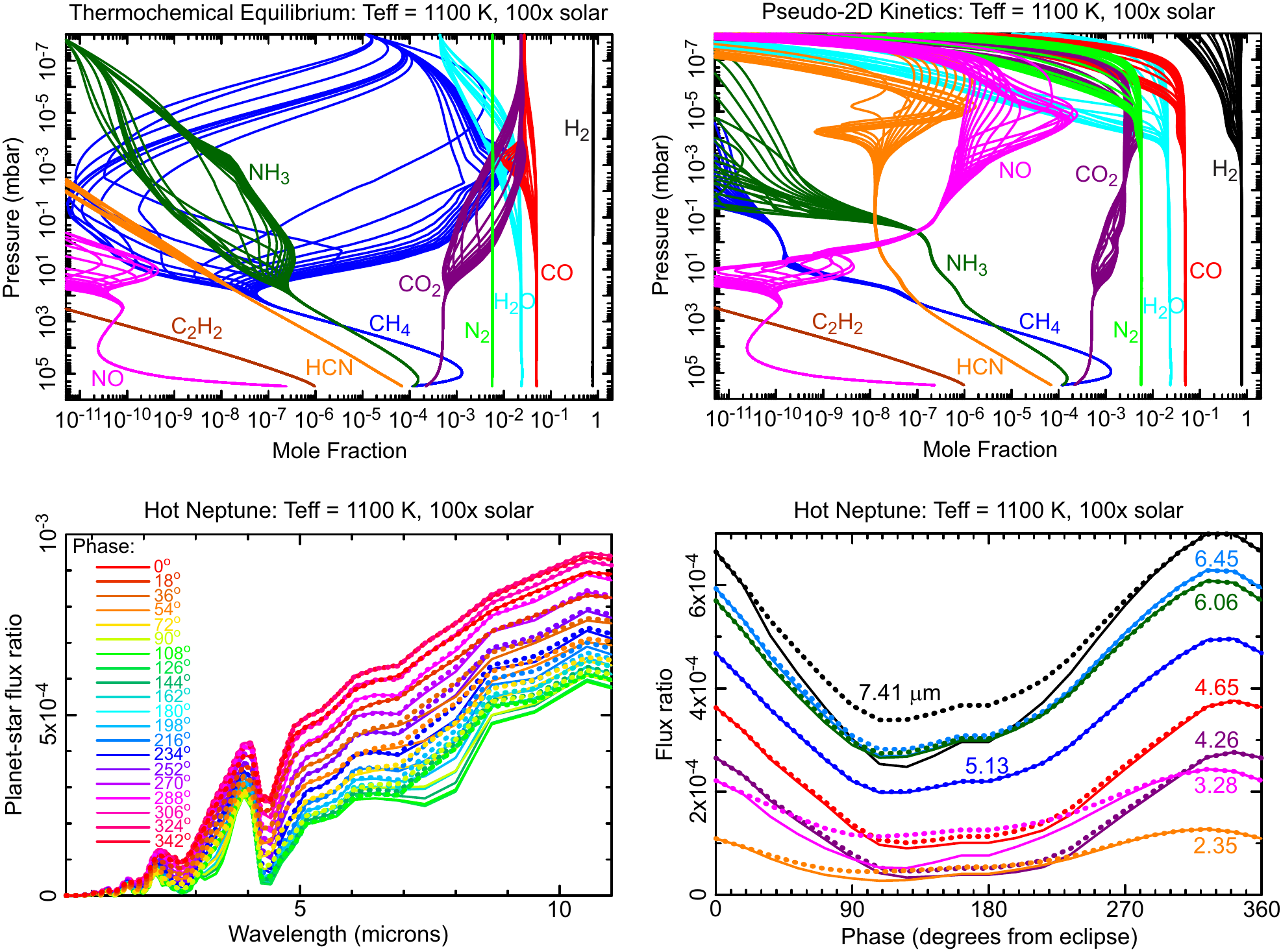} 
\end{center}
\vspace{-0.5cm}
\caption{Same as Fig.~\ref{fig700p2}, except for the \teff\ = 1100 K, 100$\times$ solar metallicity exo-Neptune.}
\label{fig1100p2}       
\end{figure*}

Figure \ref{figtemp100x} shows the 2D thermal structure of our 100$\times$ solar metallicity Neptune-class exoplanets for assumed planetary 
\teff\ = 700, 900, and 1100 K.  
Stratospheric temperatures in the $\sim$0.1-1000 mbar region are greater with the 100$\times$ solar metallicity models in comparison with 
solar-composition models for planets at the same orbital distances, except at high altitudes at some night-side longitudes, where the 
100$\times$ solar model can become cooler due to shorter radiative time constants.  Temperature variations with longitude 
and day-night temperature contrasts are also greater with the 100$\times$ solar model.  The greater abundance of heavy molecular 
constituents in the 100$\times$ solar metallicity case causes the location of the infrared photosphere to decrease in pressure \cite{lewis10,drummond18metal}, 
essentially lifting the entire temperature profile up in log-pressure space.  Temperatures are much more variable with longitude at the lower 
pressures where the emission originates in the 100$\times$ solar metallicity models compared to the 1$\times$ solar case.
These differences in thermal structure lead to large changes in atmospheric composition and emission properties compared with the 1$\times$ 
solar case (see Figs.~\ref{fig700p2}, \ref{fig900p2}, \ref{fig1100p2}).

Note also from comparisons of Figs.~\ref{figtemp} \& \ref{figtemp100x} that the CO = CH$_4$ equal-abundance curve has shifted to lower temperatures at all 
pressures when the metallicity is changed from 1$\times$ to 100$\times$ solar.  As the atmospheric metallicity increases, 
molecules with more than one heavy element become favored in thermochemical equilibrium over molecules with only one heavy element.  On the other hand, 
the equilibrium cloud condensation curves, whose analytic expressions derive from \cite{visscher10rock,morley12}, have shifted to higher temperatures 
at any particular pressure.  
These equilibrium clouds are expected to condense deeper in the atmosphere as metallicity increases.  At the \teff 's shown in Fig.~\ref{figtemp100x}, the 
magnesium silicate clouds may condense too deep to affect photospheric properties, but other clouds with non-trivial expected mass column densities such as 
Na$_2$S and KCl \cite{morley12,parmentier16} --- not considered in our pseudo-2D models --- could be present on the night side and/or day side of these planets.

For the 100$\times$ solar metallicity, \teff\ = 700 K case shown in Fig.~\ref{fig700p2}, CO is the dominant stratospheric carbon constituent in the 
pseudo-2D kinetics model at all longitudes, unlike the 1$\times$ solar model, and CO$_2$ has become several orders of magnitude more abundant.  Species 
with one heavy atom, such as H$_2$O, NH$_3$, CH$_4$, have also increased, but not as dramatically as CO and especially CO$_2$.  Horizontal quenching 
is still effective, keeping the stratospheric abundances constant with longitude, except at higher altitudes where photochemistry dominates.  HCN, C$_2$H$_2$, 
and NO are still important disequilibrium products.  For 
the thermochemical-equilibrium model, CO is the dominant carbon constituent on the day side, but CH$_4$ dominates on the night side at $P$ $\lta$ 1 mbar, 
illustrating the shift in the CO stability field with metallicity mentioned above.  When CO and CH$_4$ switch places in the dominant spot, the H$_2$O 
mixing ratio is also affected, due to the 
competition for oxygen between CO and H$_2$O. Carbon dioxide also exhibits strong day-night variability, being more abundant at night.  These 
abundance variations, along with the upward shift in the photosphere and the strong longitudinal variations in temperature at these pressures, appear 
prominently in the emission spectra and phase curves seen in Fig.~\ref{fig700p2}.  Variations in emission with planetary phase are much greater with 
the 700 K, 100$\times$ solar metallicity model than they were with the 1$\times$ solar model, and are especially pronounced with the thermochemical 
equilibrium model.  The phase-curve amplitudes are smaller with the pseudo-2D kinetics model than the equilibrium model because of the lack of strong 
longitudinal variations 
in species abundances.  The differences between the equilibrium and disequilibrium models are especially obvious in wavelength regions where CH$_4$ 
absorbs --- e.g., the bands centered near 7.8 and 3.3 $\mu$m --- but differences are also obvious in the water bands.  Note that the water band in the 
$\sim$5.5-7 $\mu$m region in the equilibrium model switches from being in absorption when the night side of the planet is in view, to being in emission 
when the day side is in view.  The same is not true of the CH$_4$ band in the $\sim$7-8 $\mu$m region --- there is insufficient CH$_4$ in the 
equilibrium model on the day side when the temperature inversions are present to have the CH$_4$ appear in emission.  In terms of \textit{Ariel} 
detectability, our models suggest that the phase curve for the 100$\times$ solar, \teff\ = 700 K planet would be detectable in the 7.8 $\mu$m 
methane band with \textit{Ariel} if the atmosphere were to remain in thermochemical equilibrium, but not detectable at shorter wavelengths 
or for the case where disequilibrium chemistry dominates.

The \teff\ = 900 K, 100$\times$ solar metallicity results have many similarities to the 700 K case (cf.~Figs.~\ref{fig700p2} \& \ref{fig900p2}).  The same four 
quenched species --- CO, H$_2$O, N$_2$, and CO$_2$ --- dominate in the stratosphere in the pseudo-2D kinetics model, but the quenched mixing 
ratios of CH$_4$ and NH$_3$ are smaller than they were in the \teff\ = 700 K case.  Photochemical production of HCN and especially C$_2$H$_2$ 
is reduced, but the NO abundance increases under these conditions.  In the thermochemical-equilibrium model, CO, CH$_4$, and H$_2$O still 
exhibit substantial diurnal/longitudinal abundance variations, but variations in CO$_2$ are not as large as they were for the 700 K case.  
The phase-curve amplitudes for the pseudo-2D \teff\ = 900 K model are larger than they were for the 700 K model, as a result of the greater 
diurnal temperature variations, and should be detectable in AIRS channel 1 for both the equilibrium and disequilibrium cases.  The emission spectra 
and phase curves are similar for the pseudo-2D kinetics and thermochemical-equilibrium 
models at most wavelengths, except where CH$_4$ absorbs.  The CH$_4$ absorption in the night-side hemisphere of the thermochemical-equilibrium model 
is much greater than that of the pseudo-2D model, and the large diurnal variations in the CH$_4$ abundance in the equilibrium model lead to 
large phase-curve amplitudes at wavelengths where CH$_4$ absorbs ($\sim$2.3, 3.3, and 7.8 $\mu$m).

Figure \ref{fig1100p2} shows that at \teff\ = 1100 K, 100$\times$ solar metallicity, the mixing-ratio profiles in the pseudo-2D kinetics model 
more closely resemble the thermochemical-equilibrium solution.  Methane still exhibits strong diurnal variations in the equilibrium model, but 
CH$_4$ only becomes a major species at relatively higher altitudes on the night side, and the phase-curve differences between the two models are therefore 
more muted than they were at the cooler temperatures.  The increased night-side absorption at CH$_4$-band wavelengths is still apparent in the 
equilibrium-model spectra, however.  Phase-curve variations should be readily detectable with \textit{Ariel} for both the equilibrium and disequilibrium 
cases.

All three 100$\times$ solar models show evidence for a wavelength-dependent offset of the maximum emission away from 0$\deg$ phase, where the 
day side of the planet is fully facing the observer.  Different wavelengths probe different depths in the atmosphere.  The phase offsets can provide 
useful information about atmospheric dynamics and other characteristics of these planets \cite{fort06a,knut09,lewis13,cowan12,schwartz17,parmentier18}.

\section{Comparisons with other multi-dimensional models} \label{sec:comparisons}

As discussed in section \ref{sec:chemresults}, vertical quenching becomes less important than zonal (i.e., longitudinal) quenching in controlling where a species 
abundance first departs from equilibrium in hotter atmospheres for which the dayside vertical quench point resides in a pressure region in which 
temperatures are notably changing with longitude.  This situation occurs in our models at planetary \teff\ $\ge$ 1500 K for the CO-CH$_4$-H$_2$O quench 
point and at planetary \teff\ $\ge$ 1900 K for the N$_2$-NH$_3$ quench point.  Previous pseudo-2D and 3D GCMs that include chemical quenching have typically 
considered hotter planets that fall into this situation of a high-altitude dayside vertical quench point and dominance of zonal quenching.  For example, 
the pseudo-2D chemical models and GCMs of the hot Jupiters HD 189733b (\teff\ = 1700 K) and HD 209458b (\teff\ = 2090 K) by Ag{\'u}ndez et al. 
\cite{agundez14pseudo} and Drummond et al. \cite{drummond18,drummond20} and the pseudo-2D chemical models and GCMs of WASP-43b (\teff\ = 1900 K) by 
Mendon{\c c}a et al. \cite{mendonca18} and Venot et al. \cite{venot20wasp43} obtain results in this higher-quench-point regime, where vertical 
quenching is apparently less important in controlling the global composition in the deeper photosphere where the abundance profiles first depart from 
equilibrium.  In all models, regardless of planetary \teff, zonal and/or meridional quenching will affect the global abundances when transport time 
scales are shorter than chemical lifetimes.

Our results are qualitatively consistent with those from previous solar-composition hot-Jupiter pseudo-2D chemistry models \cite{agundez14pseudo,agundez12,venot20wasp43} 
and 3D GCM results that include chemistry \cite{drummond18,drummond20,mendonca18}, but we do see some differences.  
For example, our dayside vertical quench points are generally 
deeper for any given \teff\ than in the Ag{\'u}ndez et al. \cite{agundez14pseudo}, Drummond et al. \cite{drummond18,drummond20}, or Mendon{\c c}a et al. 
\cite{mendonca18} models, leading to vertical quenching being more important in controlling where the abundance profiles depart from equilibrium in our 
models.  This difference arises for several reasons.  First, we use the Moses et al. \cite{moses11,moses13gj436} reaction mechanism, 
whereas most of these previous models use the Venot et al. \cite{venot12,venot19reduced} mechanism (the WASP-43b pseudo-2D study of \cite{venot20wasp43} 
being the exception).  With the Venot et al.~mechanism, key species such as CH$_4$ and NH$_3$ can remain in equilibrium to lower temperatures, resulting in a 
lower-pressure quench point \cite{venot12,moses14}.  Second, the \texttt{2D-ATMO} temperature profiles used in our study converge to a longitude-independent 
steady-state adiabatic profile at depth, where the atmosphere is convective.  Our profiles do not show significant variations in radiatively-controlled 
temperatures at pressures greater than $\sim$200 mbar for any planetary \teff.  The GCMs, on the other hand ---  or 2D models in which the temperature structures 
are derived from GCMs --- exhibit zonal temperature variations to pressures at least an order of magnitude greater than our Neptune-class \texttt{2D-ATMO} 
models.  This shallower radiative ``skin depth'' in our models may relate to the lower gravity on our Neptune-class planets in comparison with equivalent \teff\ hot 
Jupiters; essentially, the Neptune-class planets have a greater atmospheric scale height and therefore a greater column abundance available for absorption of 
the stellar radiation above any particular pressure level.  It is also possible that the temperatures at deep levels in the GCMs are affected by complicated 
dynamics not present in the \texttt{2D-ATMO} model.  The third and probably most important reason for the differences is that the GCMs are predicting very weak 
vertical transport at the lowest levels in the models, whereas the \texttt{2D-ATMO} models extend to greater depths, where the atmosphere is convective and 
vertical transport is stronger.  In our pseudo-2D models, we adopt a larger $K_{zz}$ to represent this convective region below the radiative-convective boundary 
of the \texttt{2D-ATMO} results.  

The very low effective $K_{zz}$ values at high pressures inferred from the vertical transport time scales presented in the above-mentioned hot-Jupiter GCM studies 
is somewhat surprising, even though these models do not extend down into the convective region of the atmosphere.  Vertically propagating gravity waves 
or other planetary waves --- often generated from deeper levels --- can dominate the effective vertical mixing in the radiative region of a real atmosphere, resulting 
in the classical inverse-square-root pressure dependence of $K_{zz}$ in planetary middle atmospheres \cite{lindzen81}.  Such a dependence was derived for HD 209458b 
from the GCM tracer mixing study of Parmentier et al. \cite{parmentier13}, and the inferred $K_{zz}$ values at 1 bar in the Parmentier et al. study are 
significantly greater than those in the GCMs mentioned above.  The quoted vertical transport time constants in those papers were estimates only, but it may be worth 
noting that the GCM results for pressures greater than a few bar require longer spin-up times than were considered in these simulations (e.g., \cite{parmentier13}).  
It is therefore possible that the extremely weak vertical transport at depth in the GCMs is an artifact of the short spin-up times or boundary conditions imposed 
in the GCMs (see \cite{carone20,mendonca20,wang20}) and that vertical mixing is more efficient than is indicated by the coupled chemistry-transport derived from GCMs.  
These concerns are unlikely to strongly affect predictions for hotter planets, where the composition largely remains close to thermochemical equilibrium throughout 
the photosphere, but they are worth considering for cooler planets.
On the other hand, pseudo-2D models have highly idealized transport, and in particular do not consider meridional transport or altitude-dependent zonal transport.  
Drummond et al. \cite{drummond18,drummond20} demonstrate that under certain conditions, meridional transport can strongly influence the global abundance of 
quenched species in giant-exoplanet atmospheres.  In these cases, higher abundances 
of a quenched constituent such as CH$_4$ at cooler mid-latitudes can penetrate into the lower-latitude regions, affecting the global abundances.  The pseudo-2D models 
would then under-predict the CH$_4$ abundance in such cases.  

In truth, a hierarchy of models is useful in exoplanet chemical studies, as all types of models have unique 
strengths and weaknesses and valuable insights to offer.   Complex atmospheric photochemistry is best captured in 1D or pseudo-2D models, at this point in time, 
due to the computational expense of including large chemical reaction lists in 3D coupled radiative-dynamical-chemical models.  These 1D or 2D models can also more efficiently 
explore a range of parameter space, such as with the grid of planets explored here.  However, if photochemistry is expected to have a relatively minor influence on 
observed spectra (i.e., for warmer atmospheres), then 3D GCMs that consider chemical relaxation or that include a smaller number of chemical reactions that are 
relevant to transport-induced quenching are highly desirable, in order to have the greatest confidence in the global predictions --- at least, assuming that the 
aforementioned effects of short spin-up times or shallow lower boundaries are well understood.

\section{Caveats with respect to the phase-curve predictions} \label{sec:caveats}

Several other caveats need to be mentioned in relation to this modeling and the resulting phase-curve predictions.  
First, we do not self-consistently calculate the temperature profiles for the pseudo-2D kinetics case.  Self-consistent calculations
\cite{drummond16,drummond18,drummond20,hu14,mendonca18} suggest that the feedbacks on temperature for hot Jupiters are of order 10\%, but 
that result will depend on the planetary \teff , the dominant constituents contributing to the atmospheric opacity, and whether those species are affected 
by quenching.  Second, our kinetics model does not consider condensation and so assumes that the gas-phase bulk elemental ratios are constant.  In real exoplanet 
atmospheres, condensation --- especially of a major component like MgSiO$_3$ --- can affect the gas-phase elemental ratios, depleting the elements that 
become tied up in condensates.  The gas-phase elemental ratios therefore can vary strongly below and above the condensation regions, affecting the abundances 
of key opacity sources such as H$_2$O \cite{helling16}.  We partially take this effect into account by removing $\sim$20\% of the oxygen from planets whose 
Mg-silicate clouds condense well below the photosphere, but the situation can become more complicated for planets in which the silicate clouds reside in 
the photosphere itself.
Third, our chemical-kinetics modeling considers only C-, N-, O-, and H-bearing species.  Other elements can be present in the infrared photosphere and 
might have interesting chemical kinetics that could affect the predicted phase-curve variations (e.g., \cite{wang17}).
Fourth, our pseudo-2D models do not consider winds (either vertical, meridional, or non-constant zonal winds), which can 
affect the 3D distributions of the species.  The lack of realistic 3D winds is a potentially significant limitation of the pseudo-2D modeling approach, as 
3D models that included chemical quenching demonstrate that meridional transport in some cases can be very important in affecting species distributions 
\cite{drummond18,drummond20,mendonca18}.

Finally, our 2D thermal-structure models do not consider clouds or hazes.  Clouds can certainly 
modify the thermal structure and phase-curve behavior for exo-Neptunes and other exoplanets \cite{charnay15,parmentier16}.  If present at altitudes above where 
the photosphere is expected 
for cloudless conditions, optically thick clouds can raise the effective photosphere to lower pressures, which reduces the emission if
temperatures are decreasing with altitude; molecular absorption bands can also be suppressed in this situation \cite{liou02}.  Figs.~\ref{figtemp} \&
\ref{figtemp100x} include condensation curves for a few interesting equilibrium cloud condensates, but may other cloud candidates exist in these atmospheres 
\cite{morley12,marley13,helling19rev}.  Not all clouds will be equally important for the emission behavior.  There is a lot more material available to form magnesium-silicate 
clouds such as  enstatite and forsterite, for example, than there is to form MnS, KCl, or ZnS clouds \cite{fort05}.   Based on microphysical modeling, Gao et 
al. \cite{gao20} suggest that at least in terms of transmission observations, Mg-silicate clouds dominate the aerosol opacity for solar-composition 
hot Jupiter exoplanets with \teff\ $>$ 950 K, whereas photochemically produced hydrocarbon aerosols dominate for \teff\ $<$ 950 K, and extinction from these 
two sets of aerosols can explain much of the variation in the 1.4 $\mu$m water band as function of planetary \teff\ that is observed by the WFC3 instrument 
on the \textit{Hubble Space Telescope} (HST).  In laboratory simulations, photochemically produced aerosols are generated under a variety of background 
atmospheric conditions relevant to exo-Neptunes \cite{he18size,he18,he19,horst18,fleury19,moran20}.  Clouds may not be enshrouding the entire planet.  
Depending on planetary \teff , Figs.~\ref{figtemp} \&
\ref{figtemp100x} show situations where certain cloud species would only condense at cooler longitudes on the night side of the planet.  Even if present on 
both the day side and night side, the condensation altitudes and the vertical cloud extent can be different \cite{helling20}.  The 3D distribution of clouds on 
hot Jupiters has been explored by a few groups
\cite{charnay15,helling16,helling20,lee16,lee17,lines18cloudy,lines18exonephology,lines19overcast,parmentier13,parmentier16,roman20exocloud,venot20wasp43}, and the effects of 
clouds on the thermal emission and phase-curve behavior has been investigated by Charnay et al.~\cite{charnay15}, Parmentier et al.~\cite{parmentier16}, and 
Roman et al.~\cite{roman20exocloud}.
Night-side clouds are expected to increase the phase-curve amplitude and decrease the hot-spot offset of hot Jupiters \cite{parmentier16}, and these studies so 
far suggest that clouds would not adversely affect the detection of phase-dependent emission differences, but just complicate the interpretation.

\section{Conclusions and implications for Ariel observations} \label{sec:implications}

We have developed 2D thermal structure models and pseudo-2D chemical kinetics models for Neptune-class exoplanets with a variety of planetary \teff\ to 
explore how atmospheric temperatures and composition change as a function of altitude and longitude within the equatorial regions of planets at different 
distances from their stellar host.  We also 
examine how the temperature fields and 2D composition vary as a function of atmospheric metallicity.  As with previous pseudo-2D modeling of hot Jupiters 
\cite{agundez14pseudo,agundez12,coop06,venot20wasp43}, we find that horizontal transport-induced quenching is very effective in our simulated exo-Neptune 
atmospheres, acting to homogenize the vertical profiles of species abundances with longitude, except at high altitudes in the stratosphere, where photochemical 
processes dominate.  Although we have not directly investigated the influence of planetary size or gravity, we find that our results are qualitatively 
similar to those of Jupiter-sized planets with similar effective temperatures \teff\ and metallicities.  
Our modeling suggests that horizontal quenching will be common on most planets with thick, deep atmospheres that possess equatorial zonal 
wind jets.  We find strong differences between our predicted 2D distribution of atmospheric species in our disequilibrium chemistry models compared 
with thermochemical-equilibrium predictions for exo-Neptunes with \teff\ $\le$ 1100 K.  At \teff\ $\ge$ 1300 K, however, the atmospheric temperatures are 
large enough that thermochemical equilibrium can be kinetically maintained, reducing the differences between equilibrium and disequilibrium predictions.  As 
with other models that explore trends in planetary emission with \teff\ \cite{perez13,komacek16}, we predict 
that day-night temperature contrasts and phase-curve amplitudes will increase with increasing \teff .  Actual observations, however, exhibit more complicated 
trends \cite{komacek17,parmentier18}, suggesting other processes such as globally inhomogeneous clouds --- which are not considered in our models --- 
contribute to global thermal-structure variations.  Our models have important implications for planning and prioritizing phase-curve observations with 
\textit{Ariel} and other spectroscopic missions such as the \textit{JWST}.

Our modeling suggests that cloudless, H$_2$-rich Neptune-sized exoplanets can be good candidates for phase-curve observations with \textit{Ariel}, 
depending on their effective atmospheric temperatures (i.e., distance from their host stars) and other atmospheric properties such as metallicity and chemical 
regime (i.e., thermochemical equilibrium versus effective disequilibrium chemistry).  For exo-Neptunes with solar-composition atmospheres, cooler planets with 
\teff\ = 500--700 K have interesting predicted differences in emission spectra depending on whether their atmospheres are in thermochemical equilibrium or 
whether disequilibrium processes such as transport-induced quenching and photochemistry are operating, especially at wavelengths where CO, CO$_2$, and 
NH$_3$ have molecular bands (see Fig.~\ref{figphase}).  Such planets are therefore appropriate candidates for eclipse spectroscopy, where the goal would be 
to determine the atmospheric composition, to constrain planetary-formation scenarios, and to identify the dominant chemical processes currently at play in exoplanet 
atmospheres.  However, our models predict that these cooler solar-composition planets would have relatively flat phase curves, making them unsuitable targets for 
time-intensive phase-curve observations with \textit{Ariel}.  

In contrast, we predict that solar-composition exo-Neptunes with \teff\ $\ge$ 1300 K have 
emission spectra that vary strongly with orbital phase, due to large diurnal variations in stratospheric temperatures; the phase amplitudes grow with 
increasing planetary \teff .  Molecular bands of H$_2$O and CO are expected to be readily apparent on such planets at \textit{Ariel}-sensitive wavelengths 
(see Fig.~\ref{figphase}), and phase-curve observations will provide good tests of our understanding of the thermal structure, energy balance, and 
atmospheric dynamics on such planets
\cite{charnay20,cowan11model,fort06a,kataria15,komacek16,komacek17,knut07,knut12,lewis10,parmentier18,schwartz17,showman09,stevenson14wasp43b}, making them 
good candidates for phase-curve 
observations with \textit{Ariel}.  However, compositional differences between the potential equilibrium versus disequilibrium 
chemical regimes are minor on such hot planets, and the variations in the dominant atmospheric constituents with longitude are predicted 
to be small, making hot planets less interesting targets for constraining chemical processes on exoplanets (see Figs.~\ref{figmixall}).  

Solar-composition exo-Neptunes 
in the intermediate \teff\ = 900-1100 K regime fall in a ``sweet spot'' in which the phase-curve amplitudes are non-trivial (especially at longer wavelengths), 
and the atmospheric composition is diagnostic of the equilibrium vs.~disequilibrium chemical regime.  Interesting chemical variations can occur on these planets 
over the course of the planet's orbit, at least if thermochemical equilibrium is maintained; horizontal quenching tends to homogenize the composition as a function of 
longitude in our pseudo-2D models, except at high altitudes where photochemistry dominates.  Such intermediate \teff\ planets would appear to be excellent 
targets for \textit{Ariel} phase-curve observations from the standpoint of constraining the atmospheric chemical processes along with the above-mentioned 
radiative and dynamical processes, albeit inherently riskier targets due to smaller predicted phase-curve amplitudes when disequilibrium chemistry is considered.

At higher atmospheric metallicities, which is expected to be a more realistic assumption for Neptune-sized planets \cite{fort13frame,helled14uran}, this 
``sweet spot'' shifts to smaller planetary \teff\ because the increased atmospheric opacity causes the infrared photosphere to shift lower pressures where 
temperatures vary more significantly with longitude.  As was previously shown with 3D GCM modeling \cite{lewis10,kataria15}, the increase in metallicity leads to 
an increase in phase-curve amplitudes as a result of this effect.  Therefore, even our cooler \teff\ = 700 K exo-Neptune model with an atmospheric metallicity 
of 100$\times$ solar is expected to exhibit prominent changes in emission with orbital phase (see Fig.~\ref{fig700p2}), although the phase-curve amplitude is 
much greater if the atmosphere remains in thermochemical equilibrium, compared to our disequilibrium pseudo-2D model.  As planetary \teff\ increases with 
the 100$\times$ solar metallicity models, the phase-curve amplitudes also increase, due to the larger variations in stratospheric temperatures with longitude, 
but the composition differences between the thermochemical-equilibrium and pseudo-2D kinetics models begin to decrease, as with the lower-metallicity models.  
The main difference between the equilibrium and disequilibrium models with the 100$\times$ solar metallicity models in the planetary \teff\ = 700-1100 K range 
is the abundance of CH$_4$, which exhibits very strong diurnal variations in the thermochemical-equilibrium model that are very apparent in the emission 
spectra in the broad methane bands centered at 2.3, 3.3, and 7.8 $\mu$m.  Wavelength-dependent hot spot offsets are also more apparent with higher-metallicity 
atmospheres, introducing the possibility that atmospheric dynamics and radiative properties could be constrained as a function of altitude from phase-curve 
observations.  For cool exo-Neptunes around nearby stars that are bright as seen from Earth, a higher-than-solar metallicity could first be confirmed from 
initial \textit{Ariel} transit and/or eclipse observations, followed by full phase-curve observations if models such as a presented here indicate that 
observable phase-curve amplitudes are likely.  Such planets would help constrain both physical and chemical processes in exo-Neptune atmospheres.

In summary, smaller Neptune-class planets could be excellent candidates for phase-curve observations with \textit{Ariel}, depending on planetary \teff\  
and atmospheric metallicity, 
providing important information regarding exoplanet atmospheres that cannot be acquired by any other means.  
Initial transit and eclipse observations could help identify the most promising targets and help evaluate the impact of clouds and other complicating 
factors that can affect phase-curve spectra.
For a more detailed discussion of phase-curve observations with \textit{Ariel}, see Charnay et al.~\cite{charnay20}.



\begin{acknowledgements}
We gratefully acknowledge support from the NASA Exoplanets Research Program grant NNX16AC64G (J.M.), the European Research Council Grant Agreement ATMO 757858 (P.T.), 
the CNRS/INSU Programme National de Plan{\'e}tologie (PNP) and CNES (O.V.).  
\end{acknowledgements}

%
%

\bibliographystyle{spmpsci}      
\bibliography{/data/Tex/Papers/references}   

%
%

\end{document}